\pdfoutput=1
\documentclass[twocolumn]{aastex701}

\usepackage{amsmath}
\usepackage{CJK}
\usepackage{bm}
\usepackage{hyperref}

\begin{document}
\begin{CJK*}{UTF8}{gbsn}
\title{Modeling the Evolution of Protoplanetary Disks: Two Pathways from Gravitational Instability to MHD Wind-Driven Accretion}

\author[0000-0003-0794-1949]{Yang Ni (倪阳)}
\affiliation{Institute for Advanced Study, Tsinghua University, Beijing 100084, China}
\email[show]{ny22@mails.tsinghua.edu.cn}

\author[0000-0002-9408-2857]{Wenrui Xu (许文睿)}
\affiliation{Center for Computational Astrophysics, Flatiron Institute, New York 10010, USA}
\email[show]{xuwenrui26@gmail.com}

\author[0000-0001-6906-9549]{Xue-Ning Bai (白雪宁)}

\affiliation{Institute for Advanced Study, Tsinghua University, Beijing 100084, China}
\affiliation{Department of Astronomy, Tsinghua University, Beijing 100084, China}
\email[show]{xbai@tsinghua.edu.cn}
\begin{abstract}

The global evolution of protoplanetary disks (PPDs) sets the initial conditions for planet formation. However, most models focus on individual evolutionary phases, with idealized initial conditions and oversimplified prescriptions for angular momentum transport and thermodynamics. We present a semi-two-dimensional ($1+1$D) model incorporating gravitational instability (GI), magnetohydrodynamic (MHD) winds, magneto-rotational instability (MRI), stellar irradiation, self-shadowing, and radiation transport. The radial distribution of large-scale magnetic flux drives two pathways of disk evolution. When the vertical field is spatially uniform, a puffed-up, MRI-heated inner rim shadows the disk beyond it. The shadow sustains a massive, marginally gravitationally unstable region for $\sim1\,\mathrm{Myr}$ and, for several Myr, a compact ($\lesssim10\,\mathrm{AU}$), cold ($\sim10\,\mathrm{K}$), low-turbulence ($\alpha_\mathrm{SS}\sim10^{-4}$), high-density ($\Sigma\gtrsim300\,\mathrm{g\,cm^{-2}}$), optically thick reservoir. A CO-based gas mass measurement would therefore underestimate the true disk mass by more than an order of magnitude at later times. When the field instead scales with midplane gas pressure, it drives stronger transport in the inner disk and eventually strips the shadow, leaving an extended, flared disk whose observable mass closely traces the true mass. Our results connect GI-dominated Class~0/I disks to MHD wind-driven Class~II disks, and point to three broader conclusions: (i) disk physics is strongly inhomogeneous in space and time, so constant-$\alpha$ treatments miss essential physics; (ii) thermodynamics plays an active role, with self-shadowing simultaneously preserving GI and weakening MHD winds; and (iii) the distribution of large-scale magnetic flux is the key uncertainty. The two pathways are qualitatively consistent, respectively, with observations of compact, shadowed disks and extended, irradiated disks. 

\end{abstract}


\section{Introduction} \label{sec:intro}

Protoplanetary disks (PPDs) are the primary sites for planet formation, and their global evolution plays a fundamental role in shaping the architecture and demographics of planetary systems. The surface density, temperature structure, magnetic activity, and level of turbulence in disks regulate key processes such as dust growth and radial drift, planetesimal formation, pebble accretion, and planet migration. Since disks evolve over timescales of a few million years, understanding how angular momentum transport and mass loss operate throughout their lifetimes is essential for connecting disk physics to observed exoplanet populations \citep[e.g., see][for reviews]{Armitage_2011ARA&A, Manara_2023ASPC, Miotello_2023ASPC}.
Moreover, observational evidence increasingly suggests that planet formation begins early in disk evolution. High-resolution ALMA observations have revealed that substructures such as rings and gaps are ubiquitous in Class II disks \citep[e.g.][]{Andrews_2018ApJ, Huang_2018ApJ}, and similar features have now been detected in younger, deeply embedded Class 0/I systems as well \citep[e.g.][]{Tobin_2020ApJ, Hsieh_2024ApJ}. These findings imply that the physical conditions conducive to planet formation are already established at early stages, when disks are still massive and dynamically evolving. This motivates a coherent theoretical framework capable of describing disk evolution continuously from the embedded Class 0/I phase through the more evolved Class II stage.

From a theoretical perspective, disk evolution has traditionally been modeled using simplified prescriptions for angular momentum transport. A large class of models adopts a viscous framework based on the $\alpha$-disk formalism \citep{Shakura_1973A&A, Lynden-Bell_1974MNRAS, Pringle_1981ARA&A}, with extensions incorporating gravitational instability (GI; \citealt{Lin_1987MNRAS}) or magneto-rotational instability (MRI; \citealt{Zhu_2010ApJ}). More recently, models emphasizing magneto-hydrodynamic (MHD) disk winds as the dominant driver of accretion and dispersal have been developed \citep[e.g.][]{Suzuki_2016A&A, Bai_2016ApJ, Tabone_2022MNRAS}. These viscous, wind-driven, or hybrid models have been widely applied to studies of disk evolution and planet population synthesis \citep[e.g.][]{Coleman_2014MNRAS.445..479C, Emsenhuber_2021A&A...656A..69E, Guo_2025ApJ...983...56G}. Despite their success, most existing models rely on strong simplifications that limit their physical realism. In particular, angular momentum transport is often parameterized using constant or weakly varying $\alpha$ values, implicitly assuming that disk physics is homogeneous in space and time. Such simplified prescriptions can leave strong imprints on the predicted disk structure: for example, the wind-driven model of \citet{Suzuki_2016A&A} produces a pressure bump in the inner disk, yet one should be careful in taking it as a robust prediction. Some models attempt to incorporate additional effects, such as dead zones \citep[e.g.][]{Hasegawa_2015ApJ, Tong_2025MNRAS} or layered accretion \citep[e.g.][]{Gammie_1996ApJ, Zhu_2010ApJ_2, Hasegawa_2015ApJ}, but these are typically embedded within the viscous framework and do not fully reflect our current understanding of disk gas dynamics.

Recent theoretical and numerical studies instead suggest a more complex, stage-dependent picture of disk dynamics. During the early Class~0/I phase, disks are massive and often embedded in infalling envelopes, making them susceptible to gravitational instability \citep[e.g.][]{Boss_1997Sci, Gammie_2001ApJ, Kratter_2016ARA&A, Deng_2020ApJ, Deng_2021NatAs, Xu_2021a, Xu_2021b, Bethune_2022A&A, Ni_2025ApJ...995...96N, Xu_2025}. By contrast, the role of magnetically driven disk winds in this early phase remains poorly understood. While MHD winds may in principle operate in Class 0/I disks, their efficiency is highly uncertain due to the complex thermochemical and radiative environment of embedded systems. For instance, disk geometry and self-shadowing can strongly regulate the 
penetration of ionizing stellar radiation. As demonstrated by \citet{Xu_2021b}, 
the resulting deficit in ionization within self-shadowed regions preserves strong ambipolar diffusion, which can render MHD winds weak and sub-dominant to the accretion driven by GI. Conversely, \citet{Tsukamoto_2023PASJ} demonstrate that if substantial grain growth occurs early on (to sizes $> 10\,\mu\mathrm{m}$), the reduced magnetic resistivity allows for more efficient magnetic coupling. In this scenario, the wind zone no longer suffers from strong ambipolar diffusion, and MHD wind can dominate the angular momentum transport even in the embedded phase, resulting in less massive, non-self-gravitating disks. As a result, the viability and global impact of MHD winds during the embedded phase remain an open question. 

As disks evolve into the Class~II stage, they become more optically thin following envelope dispersal, which generally allows stellar irradiation to play a more prominent role in disk heating, leading to increasingly flared and irradiated disk structures. In this regime, angular momentum transport is thought to be dominated by magnetically driven winds \citep[e.g.][]{Bai_2013ApJ}, supplemented by hydrodynamic instabilities such as the vertical shear instability (VSI; \citealt{Arlt_2004A&A, Nelson_2013MNRAS, Flock_2020ApJ, Zhang_2024ApJ}). Meanwhile, in the innermost regions where temperature and ionization levels are high, MRI-driven turbulence may still play a significant role \citep[e.g.][]{Hawley_1995ApJ, Bai_2011ApJ, Flock_2017ApJ2, Iwasaki_2024PASJ}. 

Taken together, these considerations point to a highly non-homogeneous and evolving disk environment, in which the dominant angular momentum transport mechanisms, thermal structure, and irradiation geometry vary with both time and location within the disk. Capturing the transition from massive, potentially self-shadowed and GI-dominated Class~0/I disks to flared, irradiated, wind-regulated Class~II disks therefore requires a global evolutionary model that consistently links disk thermodynamics, geometry, and magnetic activity. Motivated by this gap in our knowledge, we aim to construct such a model by incorporating the latest theoretical understanding of disk gas dynamics into a unified framework. 

In this paper, we investigate this coupled evolution by constructing a $1+1$D global model that traces the long-term evolution of PPDs from their early, potentially GI-dominated phase to their later, MHD wind-driven stage. Our model incorporates radial radiative transfer and stellar irradiation, combined with local prescriptions for GI, MRI, and MHD wind-driven angular momentum transport. This framework enables us to explore how thermal structure and magnetic field distribution jointly regulate accretion and mass loss, and to examine how different physical regimes interact and evolve throughout the disk's lifetime. The paper is organized as follows: We describe our formalism and methodology in Section \ref{sec:modelsetup}. In Section \ref{sec:results}, we study the disk evolutionary properties. We discuss our model comparison with observations as well as limitations in Section \ref{sec:discussion}, before we summarize and conclude in Section \ref{sec:conclusion}.

\section{A $1+1\mathrm{D}$ Model for PPD evolution} \label{sec:modelsetup}

To capture the disk's evolution across different epochs from the Class 0/I to Class II stage, we develop a $1+1\mathrm{D}$ model that describes the evolution of the surface density $\Sigma(R)$ and midplane temperature $T(R)$ as functions of the disk's cylindrical radius $R$. As motivated by recent studies \citep[e.g.][]{Bai_Stone_2017ApJ, Lesur_2021A&A, Xu_2021b}, disk evolution depends crucially on the radial profile of the magnetic flux distribution, as well as on the disk thermodynamics and ionization structure. We therefore highlight that our framework incorporates two models of magnetic flux evolution.

\subsection{Evolution of surface density}
\label{sec:setup_surfden}

We start by considering the vertically integrated equation for angular momentum conservation in the disk

\begin{multline}\label{eq:AMconservation}
\partial_t(\Sigma R^2 \Omega)=-\frac{1}{R} \partial_R (R^3 \Omega \Sigma v_R )\\
-\frac{1}{R} \partial_R \left(R^2 \int_{-z_\mathrm{w}}^{z_\mathrm{w}} T_{R\phi}\,\mathrm{d}z \right)-RT_{z\phi}|^{z_\mathrm{w}}_{-z_\mathrm{w}}-R^2\Omega \dot{\Sigma}_\mathrm{W},
\end{multline}
where $\Omega=\sqrt{GM_*/R^3}$ is the Keplerian angular velocity orbiting the central star of a mass $M_*$, $v_R$ is the radial velocity, $T_{R\phi} \equiv \langle\rho v_R \delta v_\phi -B_R B_\phi/4\pi \rangle$ is the time-averaged stress for radial angular momentum flux, $T_{z\phi} \equiv \langle\rho v_z \delta v_\phi -B_z B_\phi/4\pi \rangle\approx\langle-B_z B_\phi/4\pi \rangle$ is the time-averaged stress for vertical angular momentum flux, $z_\mathrm{w} \sim 4H$ is the typical height where MHD wind launches, $H=c_s/\Omega$ is the thermal scale height, $c_s=\sqrt{P/\rho}$ is the isothermal sound speed (set by disk temperature), $\dot{\Sigma}_\mathrm{W}$ is the wind-induced mass-loss per unit area per unit time, and $\delta v_\phi=v_\phi-R\Omega$ is the deviation of the rotation velocity from the Keplerian velocity.

Following the $\alpha$-disk prescription \citep{Shakura_1973A&A}, we relate the radial stress $T_{R\phi}$ to the midplane thermal pressure and parameterize $T_{R\phi}$ by 
\begin{equation}\label{eq:alpha_SS}
    \frac{2}{3}\int_{-z_\mathrm{w}}^{z_\mathrm{w}} T_{R \phi}\,\mathrm{d}z = \alpha_{\mathrm{SS}} \Sigma c_s^2,
\end{equation}
where $\alpha_{\mathrm{SS}}$ is the dimensionless Shakura-Sunyaev parameter.
We can also introduce another dimensionless $\alpha_{\mathrm{DW}}$ to parameterize the vertical stress $T_{z\phi}$ in an analogous way \citep[e.g.][]{Armitage_2013ApJ,Suzuki_2016A&A,Tabone_2022MNRAS} as
\begin{equation}\label{eq:alpha_DW}
    \alpha_{\mathrm{DW}} \equiv \frac{4R T_{\mathrm{z\phi}}|^{z_\mathrm{w}}_{-{z_\mathrm{w}}}}{3\Sigma c_s^2}\approx\frac{2R |B_z B_{\phi}|_{z=\pm z_\mathrm{w}}}{3 \pi \Sigma c^2_s}.
\end{equation}
Note that launching of MHD winds requires the presence of a large-scale poloidal (vertical) field threading the disk, and for thin disks, the mean vertical field $B_z$ is approximately constant through the disk. The value of $B_\phi$ usually scales with $B_z$ sub-linearly \citep{Bai_2016ApJ...818..152B, Lesur_2021A&A}, and the entire wind properties can be considered to be largely set by $B_z$, or the distribution of poloidal magnetic flux threading the disk. We will come back to this in Section~\ref{sec:setup_alphaDW}.

Based on Equation~(\ref{eq:alpha_DW}), we further describe the wind-induced mass-loss rate $\dot{\Sigma}_\mathrm{W}$ with the wind torque following \cite{Tabone_2022MNRAS}:
\begin{equation}\label{eq:Sigma_Wdot}
    \dot{\Sigma}_\mathrm{W}=\frac{R T_{z \phi} |^{z_\mathrm{w}}_{-z_\mathrm{w}}}{\Omega R^2 (\lambda-1)}=\frac{3\alpha_{\mathrm{DW}}\Sigma c_s^2}{4(\lambda -1)\Omega R^2},
\end{equation}
where $\lambda$ is a dimensionless parameter called wind lever arm and quantifies the ratio of specific angular momenta in the wind flow and in the Keplerian disk along the field line
\citep{Blandford_1982MNRAS,Bai_2016ApJ,Lesur_2021A&A}. For simplicity, we consider $\lambda$ as a constant both spatially and temporally, and adopt $\lambda=2$ in this study.

The equation of angular momentum conservation (\ref{eq:AMconservation}) can be combined with the vertically integrated continuity equation
\begin{equation}\label{eq:continuity}
\partial_t \Sigma=-\frac{1}{R} \partial_R(R \Sigma v_R)-\dot{\Sigma}_\mathrm{W},
\end{equation}
to yield the evolutionary equations of disk surface density $\Sigma$:
\begin{multline}\label{eq:master_surfden}
    \partial_t\Sigma=\frac{3}{R} \partial_R \left[\frac{1}{R\Omega}  \partial_R (R^2 \alpha_{\mathrm{SS}} \Sigma c_s^2) \right] \\
    +\frac{3}{2R}  \partial_R  \left( \frac{\alpha_{\mathrm{DW}} \Sigma c_s^2}{\Omega} \right)-\frac{3\alpha_{\mathrm{DW}}\Sigma c_s^2}{4(\lambda -1)\Omega R^2}.
\end{multline}
This equation is parameterized by $\alpha_{\mathrm{SS}}$,  $\alpha_{\mathrm{DW}}$ and $\lambda$ as given by Equations~(\ref{eq:alpha_SS}),~(\ref{eq:alpha_DW}), and~(\ref{eq:Sigma_Wdot}).
While this general formulation is consistent with previous 1D disk evolution models in the literature \citep[e.g.,][]{Armitage_2011ARA&A, Suzuki_2016A&A}, in those works the transport coefficients ($\alpha_{\rm SS}$, $\alpha_{\rm DW}$) and the disk temperature are typically taken to be constant or only weakly varying. In contrast, in our approach all three quantities are physically motivated and evolve self-consistently with the disk---details given in Sections~\ref{sec:setup_energyeq}--\ref{sec:setup_alphaDW}. First, we determine the disk temperature (and consequently the sound speed $c_s$) by solving the radiation transport self-consistently with the global disk structure. Second, we dynamically calculate $\alpha_{\rm SS}$ as it evolves across different spatial regions and epochs. Third, we parameterize the MHD wind torque through $\alpha_{\rm DW}$ (and potentially the mass-loading factor $\lambda$, though not explored in this work) according to the disk's global geometry, specifically accounting for the transition between stellar-irradiated and self-shadowed regions (see Section~\ref{sec:setup_windtorque}).

In principle, infall from the parent molecular cloud introduces an additional source of both mass and angular momentum that is supposed to be integrated into Equation~(\ref{eq:master_surfden}). However, we argue that for the purposes of this study, the omission of an early infall term does not qualitatively change the physical picture. By initializing our disk with a massive, marginally unstable profile, we effectively capture the state of the disk at the end of the main protostellar accretion phase. Furthermore, while recent observations suggest that ``streamers'' and late-time infall are relatively common even in the Class II stage \citep[e.g.,][]{Akiyama_2019AJ....157..165A, Gupta_2023A&A...670L...8G}, the detailed interaction between such external material and the disk's internal physical process remains poorly understood. We therefore neglect these late-stage accretion events to maintain a cleaner setup, leaving the complex interplay between streamers and e.g. disk winds for future investigation.

\subsection{Temperature Profile and Radiation Transport}
\label{sec:setup_energyeq}

With the viscous accretion and wind torque now parameterized in terms of the midplane thermal pressure, we turn to develop the evolution of the midplane temperature profile to capture a realistic representation of midplane thermal pressure and scale height throughout the PPD's evolution. The midplane temperature can be derived from the equation of energy conservation
\begin{multline}\label{eq:energy_eq}
    C_\mathrm{V} \Sigma \frac{\partial T}{\partial t}+ C_\mathrm{V} T \frac{\partial \Sigma}{\partial t} + \frac{1}{R}\partial_R(RC_\mathrm{V} \Sigma Tv_R) \\ + (\gamma-1)C_\mathrm{V} \Sigma T \frac{1}{R}\partial_R(Rv_R)
    =Q_{\mathrm{vis}}+Q_{\mathrm{irr}}-\Lambda_\mathrm{cool} \\ +\kappa \Sigma c(E_r-aT^4) +  C_\mathrm{V} T \dot{\Sigma}_\mathrm{W} ,
\end{multline}
where $C_\mathrm{V}=k_\mathrm{B}/[ (\gamma - 1) \mu m_\mathrm{p}]$ is the heat capacity, $\kappa$ is the gray opacity (and we do not distinguish the Rosseland mean and Planck mean here for simplicity), $c$ is the speed of light, $E_r$ is the radiation energy density, $a=4\sigma_{\mathrm{SB}}/c$ with $\sigma_{\mathrm{SB}}$ being the Stefan-Boltzmann constant. In this work, we adopt $\gamma=1.4$ and $\mu=2.33$. Here we use the gray atmosphere approximation and assume a simple opacity profile \citep{Semenov_2003A&A, Xu_2022ApJ}: for $T < 100\,\mathrm{K}$ we take $\kappa = \kappa_0\,(T/100\,\mathrm{K})^{2}$, while for $T \geq 100\,\mathrm{K}$
\begin{equation}
\ln \kappa = \frac{1}{2}\left[ \ln(\kappa_0 \kappa_\mathrm{sub}) + \ln\!\left(\frac{\kappa_0}{\kappa_\mathrm{sub}}\right) \tanh\!\left(\frac{T_\mathrm{sub}-T}{\Delta T_\mathrm{trans}}\right) \right],
\end{equation}
where $\kappa_0= 1\,\mathrm{cm}^{2}\,\mathrm{g}^{-1}$, $\kappa_\mathrm{sub}= 1\times10^{-3}\,\mathrm{cm}^{2}\,\mathrm{g}^{-1}$, $T_\mathrm{sub} = 1500 \,\mathrm{K}$, and $\Delta T_\mathrm{trans} = 250 \,\mathrm{K}$. 
For our purpose, the details of the opacity law do not qualitatively affect the results.

On the right-hand side of Equation~(\ref{eq:energy_eq}), the first term gives the local viscous heating \citep[e.g.][]{Bell_1994ApJ,Zhu_2010ApJ}
\begin{equation}
    Q_{\mathrm{vis}}=\frac{9}{4} \alpha_{\mathrm{SS}} \Sigma \Omega c_s^2.
\end{equation}
The second term represents irradiation heating from the center (proto-)star. To account for the self-consistent irradiation heating in our global model, we extend our model to $1+1$ D, and consider $N_\mathrm{\theta}$ meshes in the polar direction. We only utilize the polar direction for irradiation, and simply adopt a Gaussian distribution $ \rho(R,\theta)=\rho(R,0)\exp[-(R\tan\theta)^2/2H^2] $ for the vertical density. Then we compute the optical depth from $(R,\theta)$ to the central star
\begin{equation}
\tau_{*} (R,\theta)=\int_{R_\mathrm{inner}}^{R} \kappa \rho(x,\theta)\,\mathrm{d}x ,
\end{equation}
and the differential optical depth in each $\Delta R$ bin around $(R,\theta)$
\begin{equation}
\Delta\tau_{*} (R,\theta)= \kappa \rho(R,\theta) \Delta R / \cos(\theta).
\end{equation}
Now 
\begin{multline}
Q_\mathrm{irr}(R) = \frac{1}{\pi R\Delta R}\sum_{j} F_\mathrm{irr}(R,\theta_j) \Delta S(R,\theta_j) \cos(\theta_j) \\\cdot \{ 1-\exp[-\Delta\tau_*(R,\theta_j)] \},
\end{multline}
where the irradiation flux at $(R,\theta_j)$ is
\begin{equation}
F_\mathrm{irr}(R,\theta_j)=\frac{L_\odot}{4\pi\left(\frac{R}{\cos(\theta_j)}\right)^{2}}\exp(-\tau_{*}(R,\theta_j)),
\end{equation}
and the effective area receiving stellar photons at $(R,\theta_j)$ is
\begin{equation}
    \Delta S(R,\theta_j)=2\pi R^2 \frac{\Delta \theta_j}{\cos^2(\theta_j)}.
\end{equation}
Here, $j$ is the index denoting the $j$-th mesh in the polar direction.
The third term is the radiative cooling through the vertical direction as \citep{Zhu_2015ApJ, Xu_2022ApJ}
\begin{equation}
    \Lambda_\mathrm{cool}=\frac{16}{3}\sigma_{\mathrm{SB}}T^4\frac{\tau_\mathrm{mid}}{1+\tau_\mathrm{mid}^2},
\end{equation}
where the midplane optical depth is $\tau_\mathrm{mid}=\kappa \Sigma /2$.

We note that while we have accounted for radiation transport in the vertical direction, there can be additional radiative diffusion along the radial direction, which also affects the energy balance. Therefore, we consider additional radiation transport along the radial direction, together with its energy exchange with the gas, reflected in the last term in Equation~(\ref{eq:energy_eq}).
For the radiation field,
we adopt the flux-limited diffusion (FLD) approximation \citep{Levermore_1981ApJ} in the radial direction
\begin{equation}\label{eq:radialRT}
    \partial_t E_r -\nabla \cdot \left(\frac{c\lambda_R}{\kappa \rho_{\mathrm{RT}}} \nabla E_r\right) = \kappa \rho_{\mathrm{mid}} c (aT^4-E_r) ,
\end{equation}
where $\lambda_R=(2+\mathcal{R})/(6+3\mathcal{R}+\mathcal{R}^2)$ with $\mathcal{R}=|\nabla E_r|/(\kappa \rho_{\mathrm{RT}} E_r)$. In principle, $\rho_{\mathrm{RT}}$ should represent certain weighted averages along the vertical column. We take $\rho_{\mathrm{RT}}=\rho_{\rm mid}\exp(-z_{\rm RT}^2/2H^2)$, where $\rho_{\rm mid}\approx\Sigma/(\sqrt{2\pi}H)$ is the midplane gas density, $z_{\rm RT}$ is some approximate height where radial radiation transport is the most efficient. Without rigorous proof, we take $z_{\rm RT}=\min(2H, z_{\tau=1})$, where $z_{\tau=1}$ is the height of the $\tau=1$ surface, and should be taken to zero if $\kappa\Sigma/2<1$.\footnote{In the optically-thick limit, radial radiative diffusion is more efficient towards disk surface than in the midplane, but signature of radiative diffusion at the surface layer can be easily lost through surface cooling. The value of $z_{\rm RT}\sim2H$ is chosen to fit in between.} Introducing this radiative diffusion along the radial direction helps smooth the radial temperature profile, especially avoiding sharp temperature changes when the disk transitions between the irradiative and self-shadowing regimes. Similar treatments of midplane radiation transport have been implemented in 2D hydrodynamical simulations \citep[e.g.,][]{Ziampras_2026A&A}. We have verified that holding $z_{\rm RT}$ fixed at $0$, $1H$, and $2H$ leaves the sequence of evolutionary phases unchanged while shifting the evolution timescale: less efficient radial transport preserves the self-shadowed structure for longer and thereby prolongs the disk lifetime.

\subsection{Determining $\alpha_{\mathrm{SS}}$}
\label{sec:setup_alpha}

Radial transport of disk angular momentum, characterized by $\alpha_{\mathrm{SS}}$, can be mediated by various sources of disk turbulence together with some laminar magnetic stress associated with the MHD wind. Contribution from such terms is usually additive, and hence we write
\begin{equation}\label{eq:alpha_SS_Q}
    \alpha_{\mathrm{SS}}=\alpha_\mathrm{GI} + \alpha_\mathrm{WR}+ \alpha_\mathrm{MRI} + \alpha_0,
\end{equation}
and elaborate on these terms below.

PPDs in their early stages likely experience the gravitational instability \citep[GI; see][for a review]{Kratter_2016ARA&A}, which drives gravito-turbulence and transport angular momentum outward \citep{Gammie_2001ApJ}.
Although the gravitational stress in principle contains long-range contributions, numerical experiments have established that angular momentum transport in gravito-turbulent disks is reasonably well described by a local, viscous-like prescription, and that global wave transport becomes important only at large disk-to-star mass ratios \citep{Lodato_2004MNRAS, Lodato_2005MNRAS}. Recent global simulations of GI-regulated protostellar disks with full radiation transport further confirm the predominantly local nature of the transport \citep{Xu_2025}. In our calculations, long-range torques could enhance the transport during the brief early phase when the disk is most massive ($M_{\rm disk}\approx0.5\,M_\odot$ initially). Their omission is expected to mainly affect the duration of this early phase, while the qualitative evolutionary pathways, which are governed by the magnetic flux distribution, should be unaffected. We therefore adopt a local parametric prescription for GI-driven transport,
\begin{equation}
    \alpha_\mathrm{GI} \sim e^{-Q^4},
\end{equation}
where
\begin{equation}
Q=\frac{c_s \kappa_{\mathrm{ep}}}{\pi G \Sigma} \sim \frac{c_s \Omega}{\pi G \Sigma}
\end{equation}
is the Toomre Q parameter, and $\kappa_{\mathrm{ep}}\approx\Omega$ is the epicyclic frequency. This functional form serves as a phenomenological switch that encodes the self-regulated behavior of gravito-turbulence established by numerical simulations, in which the disk adjusts its transport efficiency to remain near marginal instability \citep{Gammie_2001ApJ, Kratter_2016ARA&A}. In one-dimensional models, any prescription in which $\alpha_{\rm GI}$ rises steeply as $Q$ approaches unity reproduces this self-regulation. The exponential form above has been adopted in a number of long-term disk evolution studies \citep[e.g.][]{Zhu_2010ApJ,Takahashi_2013ApJ,Kimura_2021ApJ,Xu_2025}. With this prescription, $\alpha_{\rm GI}$ is of order unity deep in the unstable regime ($Q\ll1$), declines steeply through $Q\sim1$, and is negligible in the stable regime ($Q\gtrsim2$).

The contribution from magnetic stresses is denoted as $\alpha_{\rm WR}$. This stress may result from the MRI, and/or is accompanied by the disk wind. As $\alpha_{\rm WR}\propto \langle B_RB_\phi\rangle H$, while $\alpha_{\rm DW}\propto R|B_zB_\phi|_{z=\pm z_w}$, if we consider $\langle |B_\phi|\rangle\sim |B_\phi|_{z=\pm z_w}$, and $\langle |B_R|\rangle\sim |B_z|_{z=\pm z_w}$, then it is naturally expected that $\alpha_{\rm WR}\approx(H/R)\alpha_{\rm DW}$ \citep{Wardle_2007Ap&SS.311...35W, Bai_2009ApJ...701..737B}. Such assumptions can be justified as we expect the MRI to be damped or suppressed by non-ideal MHD effects \citep{Bai_2011ApJ...736..144B, Cui_2021MNRAS.507.1106C, Xu_2021b}, thus magnetic field in the disk interior does not undergo substantial amplification compared to the field at the disk surface. As a result, we can impose
\begin{equation}
    \alpha_\mathrm{WR} = (H/R)\,\alpha_\mathrm{DW}.
\end{equation}

We further consider the innermost region of the disk, where MRI is active because of the high temperature. At $T\gtrsim10^{3}\,\mathrm{K}$, collisional ionization of alkali metals provides a sufficient ionization fraction to couple the gas to the magnetic field and activate the MRI \citep[e.g.][]{Desch_2015ApJ}. To capture this thermal-ionization transition, we follow the radiation-hydrodynamic inner-rim models of \citet{Flock_2016ApJ} and adopt
\begin{equation}
    \alpha_\mathrm{MRI} = 0.03\times\left[\frac{1-\tanh(\frac{T_\mathrm{MRI}-T}{\Delta T})}{2}\right],
\end{equation}
where $T_\mathrm{MRI}=1000\,\mathrm{K}$, and $\Delta T=25\,\mathrm{K}$.
This prescription should be interpreted as a thermal-ionization switch, not as a universal MRI saturation law. In principle, the MRI stress also depends on the net vertical magnetic flux and can reach order unity for strongly magnetized disks with $\beta_z\sim100$ \citep[e.g.,][]{2013ApJ...767...30B}. In our models, however, the regions where the thermal MRI is activated typically have $\beta_z\gtrsim10^4$, where $\alpha_{\rm MRI}=0.03$ is appropriate as a representative value.

Additionally, we set $\alpha_0 = 10^{-4}$, which represents an ``effective viscosity floor". It accounts for some base level of turbulence in the disk from other mechanisms of angular momentum transport, such as the VSI.

\subsection{Determining $\alpha_{\mathrm{DW}}$}
\label{sec:setup_alphaDW}

To estimate the wind torque, from Equation~(\ref{eq:alpha_DW}), it suffices to estimate $B_zB_\phi$ at the wind base. As stated earlier, there are two aspects involved. One is to estimate the radial profile of $B_z$ (the magnetic-flux transport; Section~\ref{sec:setup_fluxtransport}), and the other is the scaling relation between $B_\phi$ and $B_z$ (the wind-launching efficiency; Section~\ref{sec:setup_windtorque}).

\subsubsection{Prescription of Magnetic Flux Transport}
\label{sec:setup_fluxtransport}
However, the radial profile of $B_z$ and its temporal evolution are largely unconstrained observationally. Here, we prescribe this $B_z$ profile based on theoretical considerations. First, recent MHD simulations of disk formation \citep[e.g.][]{Xu_2021a, Xu_2021b, Mauxion_2024A&A} suggested that the vertical magnetic field strength $B_z$ at the Class 0/I stage can be approximated as uniform across the disk as a result of efficient radial diffusion of magnetic flux in the disk (ambipolar diffusion timescale $R^2/\eta_{\rm AD}$ is much smaller than accretion timescale $|R/v_R|$), with typical strengths of order $0.01\,$G found in those calculations. The near-uniformity of $B_z$ is a robust outcome of efficient flux diffusion. The absolute field strength, however, depends on the core-scale magnetization and other initial conditions adopted in those simulations, and remains poorly constrained. Here, we adopt $B_z=0.01\,$G as our fiducial value. Second, at the Class II stage, simulations of magnetic flux transport suggested that the disk loses magnetic flux at a rate that is faster with increasing $B_z$ \citep{Bai_Stone_2017ApJ,Lesur_2021A&A}, or decreasing plasma $\beta$ (ratio of gas to magnetic pressure). When $\beta_z\lesssim10-100$, the disk would quickly lose magnetic flux and reduce $B_z$ to maintain some reasonably large $\beta_z$. Guided by these considerations, we adopt two limiting prescriptions for $B_z$ that bracket the unknown radial flux distribution:
\begin{itemize}
\item \textbf{Uniform $B_z$} (fiducial; model \textbf{Bz\_10mG}): a spatially uniform vertical field, capped so that $\beta_z$ stays above a floor $\beta_{z,\mathrm{min}}$,
\begin{equation}\label{eq:bz}
B_z = \min\left(0.01\,\mathrm{G}, \sqrt{\frac{8 \pi \rho_{\mathrm{mid}} c_s^2}{\beta_{z,\mathrm{min}}}}\right),
\end{equation}
with $\beta_{z,\mathrm{min}} = 100$ by default and $\rho_{\mathrm{mid}}$ the midplane density. Because the gas pressure declines outward, this field becomes dynamically stronger (smaller $\beta_z$) at large radii.
\item \textbf{Pressure-scaled $B_z$} (model \textbf{Betaz\_3e4}): a field that scales with the local midplane gas pressure, so that $\beta_z$ is fixed at all radii,
\begin{equation}\label{eq:bz2}
B_z = \sqrt{\frac{8 \pi \rho_{\mathrm{mid}} c_s^2}{\beta_{z}}},
\end{equation}
with $\beta_z = 3\times10^4$ by default. Here the field is strongest where the gas is densest (the inner disk) and weakest in the diffuse outer disk.
\end{itemize}

We stress that our choice here is by no means realistic and only reflects our educated guess by combining two limiting cases. Since the $B_z$ profile is crucial in setting the overall disk evolution under wind-driven accretion, we do not consider our model to be accurate in this regard. Our focus here is the role of thermodynamics associated with the transition from the self-shadowing regime to the irradiative regime, and the new physics we incorporated in the remaining parts of our model description are much more crucial. The two primary models built on these two prescriptions are compared in Section~\ref{sec:results}, and further variations of the magnetic field configuration are explored in Section~\ref{sec:discussion}.

\subsubsection{Prescription of the Wind Torque}
\label{sec:setup_windtorque}

With $B_z$ prescribed above, evaluating the wind torque (Equation~\ref{eq:alpha_DW}) still requires the toroidal field at the wind base, i.e.\ the ratio $B_\phi/B_z$. Our prescription is motivated by two observations from recent simulations: (1) when the gas in the wind zone is well coupled to the magnetic field---as in the far-ultraviolet (FUV) ionized surface layers of directly irradiated Class~II disks---$B_\phi/B_z$ can reach as large as $\sim10$ and the wind torque is strong \citep[e.g.][]{Bai_2017ApJ, Lesur_2021A&A}; (2) in the absence of FUV penetration---as in the self-shadowed interiors of embedded Class~0/I disks---the wind zone is poorly coupled to the magnetic field with strong ambipolar diffusion, and $B_\phi/B_z$ stays small with a suppressed torque \citep[e.g.][]{Xu_2021b}. Because stronger FUV irradiation leads to better magnetic coupling, we posit that $B_\phi/B_z$ increases as the FUV ionization front reaches deeper into the disk (i.e.\ as the ionization height $z_\mathrm{FUV}$, defined below, decreases).

In the disk surface layers, FUV photons can produce an ionization fraction of $\sim10^{-5}$-$10^{-4}$ \citep{Perez-Becker_2011ApJ}, enabling strong magnetic field-gas coupling. By contrast, X-ray ionization typically yields weaker coupling at comparable heights \citep[e.g.][]{Bai_2011ApJ}. To capture the effect of FUV ionization without modeling detailed microphysics, we adopt a characteristic FUV penetration depth $\Sigma_\mathrm{FUV} = 0.01\,\mathrm{g\,cm^{-2}}$, taken to be constant with radius \citep{Perez-Becker_2011ApJ}. At each radial location $R_i$, we define an FUV ionization height $z_\mathrm{FUV}$ as the altitude above the midplane where the integrated column density along a slanted path equals $\Sigma_\mathrm{FUV}$. Specifically, we solve for the angle $\theta_\mathrm{FUV}$ satisfying
\begin{equation}
\int_{R_\mathrm{inner}}^{R_i} \rho(x, \theta_\mathrm{FUV})\,\mathrm{d}x = \Sigma_\mathrm{FUV},
\end{equation}
and compute $z_\mathrm{FUV} = R_i \tan(\theta_\mathrm{FUV})$. We implement the dependence of $B_\phi/B_z$ on $z_\mathrm{FUV}$ with a clipped interpolation in the dimensionless height $z_\mathrm{FUV}/H$:
\begin{align}
\log\left(\frac{B_{\phi}}{B_z}\right) &=
(1-f_{\rm FUV})\log\left(\left.\frac{B_{\phi}}{B_z}\right|_\mathrm{max}\right) \nonumber\\
&\quad + f_{\rm FUV}\log\left(\left.\frac{B_{\phi}}{B_z}\right|_\mathrm{min}\right), \label{eq:bphi}\\
f_{\rm FUV} &= \frac{\tilde z_\mathrm{FUV}-5H}{5H}, \\
\tilde z_\mathrm{FUV} &= \min\left[\max(z_\mathrm{FUV},5H),10H\right].
\end{align}
Here $\left.B_{\phi}/B_z\right|_\mathrm{max} = 1.9\beta_z^{0.22}$ \citep[e.g.][]{Lesur_2021A&A}, and $\left.B_{\phi}/B_z\right|_\mathrm{min} = 0.1$. Correspondingly, $z_\mathrm{FUV}\le5H$ gives the well-coupled limit, $z_\mathrm{FUV}\ge10H$ gives the weakly coupled limit, and intermediate heights are interpolated linearly in log space.

\subsection{Initial Condition and Calculation Procedures}\label{sec:method_numericalsolver}

We initialize the disk using simple piecewise power-law profiles intended to represent an early protoplanetary disk whose bulk is marginally gravitationally unstable. The model is constructed such that the region between $5\,\mathrm{AU}$ and $100\,\mathrm{AU}$ sits near the threshold for gravitational instability (GI). Within this range, we adopt radial scalings of $\Sigma \propto R^{-2}$ and $T \propto R^{-1}$, which ensures a spatially constant Toomre $Q$ parameter, here normalized to $Q = 1.2$. This setup produces a massive outer disk consistent with the expected profiles of GI-regulated Class 0/I disks \citep[e.g.][]{Xu_2021b, Mauxion_2024A&A}. The temperature profile in our initial condition is set such that this bulk GI-dominated disk is largely self-shadowing, and therefore stellar irradiation is avoided. While direct observational constraints on shadowing in embedded disks are not yet available, modeling of (sub)millimeter continuum observations of several embedded disks finds warm, optically thick, and likely gravitationally self-regulated structures whose temperatures exceed those expected from stellar irradiation alone \citep{Zamponi_2021MNRAS,Xu_2022ApJ, Xu_etal_2023ApJ}. These structures are consistent with self-shadowing, as are the disks formed in the recent high-resolution MHD simulations \citep{Xu_2021b}. Since there is little evidence that GI extends into the innermost disk, we adopt a shallower profile of surface density for $R \le 5\,\mathrm{AU}$ ($\Sigma \propto R^{-1}$) and a more flared profile of temperature ($T \propto R^{-3/4}$), and ensure the continuity of both surface density and temperature at the $5\,\mathrm{AU}$ interface. In practice, the exact initial structure of the innermost active region is not important: In our numerical experiments the MRI rapidly readjusts the surface density and temperature profiles there on a timescale less than $10^{-2}\,\mathrm{Myr}$. Beyond $100\,\mathrm{AU}$, the surface density is tapered using an exponential cutoff $\Sigma \propto R^{-2} \exp(-R/100\,\mathrm{AU})$, while the temperature is maintained at a floor of $10\,\mathrm{K}$. The total computational domain extends to $1000\,\mathrm{AU}$ to accommodate subsequent disk spreading and mass transport.

We calculate the disk profile on a log-uniform grid with $100$ mesh points extending from $0.1\,\mathrm{AU}$ to $1000\,\mathrm{AU}$ in $R$-direction, and construct a linear-uniform grid with $100$ mesh points in the $\theta$-direction for ray-tracing irradiation and deriving FUV ionization height $z_\mathrm{FUV}$. The $\theta$ domain extends from $0$ to $\pi/2$. We have tested with the $(200,200)$ resolution and found consistent results.
The inner boundary is set with zero gradients for both surface density $\Sigma$ and midplane temperature $T$. At the outer boundary, the surface density is fixed at $\Sigma_{\mathrm{min}}=10^{-6}\,\mathrm{g\,cm^{-2}}$ and the midplane temperature is fixed at $T_{\mathrm{min}}=10\,\mathrm{K}$. These values serve as numerical floors for surface density and midplane temperature, respectively.

Equations~(\ref{eq:master_surfden}),~(\ref{eq:energy_eq}),~and~(\ref{eq:radialRT}) are numerically solved using the finite difference method. The advection term in Equation~(\ref{eq:master_surfden}) is integrated with the standard upwind method, and the diffusion term is treated implicitly. Equations~(\ref{eq:energy_eq}) and~(\ref{eq:radialRT}) are coupled and solved using the backward Euler scheme \citep{Kolb_2013A&A}. Table~\ref{tab:prescriptions} summarizes the components of the model, the functional forms adopted for each, and the key parameters.

\begin{table*}[ht!]
\centering
\scriptsize
\caption{Summary of the model components, the functional forms adopted, and the key parameters. Sections~\ref{sec:setup_surfden}--\ref{sec:method_numericalsolver} give the derivations and references for each component.}
\label{tab:prescriptions}
\begin{tabular}{lll}
\hline\hline
\parbox[t]{3.0cm}{Component} & \parbox[t]{8.2cm}{Functional form} & \parbox[t]{5.2cm}{Key parameters} \\
\hline
\parbox[t]{3.0cm}{\raggedright Surface density evolution\vspace{3pt}} & \parbox[t]{8.2cm}{\raggedright Equation~(\ref{eq:master_surfden}): viscous transport with $\alpha_\mathrm{SS}=\alpha_\mathrm{GI}+\alpha_\mathrm{WR}+\alpha_\mathrm{MRI}+\alpha_0$, wind torque with $\alpha_\mathrm{DW}$, and wind mass loss $\dot\Sigma_\mathrm{W}$ (Equation~\ref{eq:Sigma_Wdot})\vspace{3pt}} & \parbox[t]{5.2cm}{\raggedright $\alpha_0=10^{-4}$; lever arm $\lambda=2$\vspace{3pt}} \\
\parbox[t]{3.0cm}{\raggedright Gravitational instability\vspace{3pt}} & \parbox[t]{8.2cm}{\raggedright $\alpha_\mathrm{GI}=e^{-Q^4}$, with $Q=c_s\Omega/(\pi G\Sigma)$\vspace{3pt}} & \parbox[t]{5.2cm}{\raggedright ---\vspace{3pt}} \\
\parbox[t]{3.0cm}{\raggedright Thermal MRI\vspace{3pt}} & \parbox[t]{8.2cm}{\raggedright $\alpha_\mathrm{MRI}=0.03\,[1-\tanh((T_\mathrm{MRI}-T)/\Delta T)]/2$\vspace{3pt}} & \parbox[t]{5.2cm}{\raggedright $T_\mathrm{MRI}=1000\,\mathrm{K}$, $\Delta T=25\,\mathrm{K}$\vspace{3pt}} \\
\parbox[t]{3.0cm}{\raggedright Wind-driven radial stress\vspace{3pt}} & \parbox[t]{8.2cm}{\raggedright $\alpha_\mathrm{WR}=(H/R)\,\alpha_\mathrm{DW}$\vspace{3pt}} & \parbox[t]{5.2cm}{\raggedright ---\vspace{3pt}} \\
\parbox[t]{3.0cm}{\raggedright Wind torque\vspace{3pt}} & \parbox[t]{8.2cm}{\raggedright $\alpha_\mathrm{DW}=2R|B_zB_\phi|/(3\pi\Sigma c_s^2)$; $B_\phi/B_z$ interpolated in log space between $1.9\beta_z^{0.22}$ at $z_\mathrm{FUV}\le5H$ and $0.1$ at $z_\mathrm{FUV}\ge10H$, with $z_\mathrm{FUV}$ set by the slanted column $\Sigma_\mathrm{FUV}$\vspace{3pt}} & \parbox[t]{5.2cm}{\raggedright $\Sigma_\mathrm{FUV}=0.01\,\mathrm{g\,cm^{-2}}$\vspace{3pt}} \\
\parbox[t]{3.0cm}{\raggedright Vertical magnetic field\vspace{3pt}} & \parbox[t]{8.2cm}{\raggedright \textbf{Bz\_10mG}: $B_z=\min(0.01\,\mathrm{G},\sqrt{8\pi\rho_\mathrm{mid}c_s^2/\beta_{z,\mathrm{min}}})$; \textbf{Betaz\_3e4}: $B_z=\sqrt{8\pi\rho_\mathrm{mid}c_s^2/\beta_z}$\vspace{3pt}} & \parbox[t]{5.2cm}{\raggedright $\beta_{z,\mathrm{min}}=100$; $\beta_z=3\times10^4$\vspace{3pt}} \\
\parbox[t]{3.0cm}{\raggedright Energy equation\vspace{3pt}} & \parbox[t]{8.2cm}{\raggedright Viscous heating $Q_\mathrm{vis}=\tfrac{9}{4}\alpha_\mathrm{SS}\Sigma\Omega c_s^2$, stellar irradiation by polar ray tracing, vertical cooling $\Lambda_\mathrm{cool}$, and radial radiative diffusion (Equation~\ref{eq:radialRT})\vspace{3pt}} & \parbox[t]{5.2cm}{\raggedright $\gamma=1.4$, $\mu=2.33$; $L_*=L_\odot$, $M_*=1\,M_\odot$; $T_\mathrm{min}=10\,\mathrm{K}$; $N_\theta=100$ over $[0,\pi/2]$\vspace{3pt}} \\
\parbox[t]{3.0cm}{\raggedright Opacity\vspace{3pt}} & \parbox[t]{8.2cm}{\raggedright $\kappa=\kappa_0(T/100\,\mathrm{K})^2$ for $T<100\,\mathrm{K}$ and $\kappa_0$ above, with a smooth transition to $\kappa_\mathrm{sub}$ across $T_\mathrm{sub}$\vspace{3pt}} & \parbox[t]{5.2cm}{\raggedright $\kappa_0=1$, $\kappa_\mathrm{sub}=10^{-3}\,\mathrm{cm^2\,g^{-1}}$; $T_\mathrm{sub}=1500\,\mathrm{K}$, $\Delta T_\mathrm{trans}=250\,\mathrm{K}$\vspace{3pt}} \\
\parbox[t]{3.0cm}{\raggedright Radial radiation transport\vspace{3pt}} & \parbox[t]{8.2cm}{\raggedright $\rho_\mathrm{RT}=\rho_\mathrm{mid}\exp(-z_\mathrm{RT}^2/2H^2)$ in the flux-limited diffusion term\vspace{3pt}} & \parbox[t]{5.2cm}{\raggedright $z_\mathrm{RT}=\min(2H,z_{\tau=1})$, and $0$ for $\tau_\mathrm{mid}<1$\vspace{3pt}} \\
\parbox[t]{3.0cm}{\raggedright Initial condition\vspace{3pt}} & \parbox[t]{8.2cm}{\raggedright $\Sigma\propto R^{-2}$ and $T\propto R^{-1}$ (i.e., constant $Q$) for $5$--$100\,\mathrm{AU}$; $\Sigma\propto R^{-1}$ and $T\propto R^{-3/4}$ inside $5\,\mathrm{AU}$; exponential cutoff beyond $100\,\mathrm{AU}$\vspace{3pt}} & \parbox[t]{5.2cm}{\raggedright initially $Q=1.2$ for $5$--$100\,\mathrm{AU}$\vspace{3pt}} \\
\parbox[t]{3.0cm}{\raggedright Grid and boundaries\vspace{3pt}} & \parbox[t]{8.2cm}{\raggedright Log-uniform radial grid; zero-gradient inner boundary; fixed outer floors\vspace{3pt}} & \parbox[t]{5.2cm}{\raggedright $100$ cells over $0.1$--$1000\,\mathrm{AU}$; $\Sigma_\mathrm{min}=10^{-6}\,\mathrm{g\,cm^{-2}}$, $T_\mathrm{min}=10\,\mathrm{K}$\vspace{3pt}} \\
\hline
\end{tabular}
\end{table*}

\section{GLOBAL DISK EVOLUTION}
\label{sec:results}

In this section, we explore the long-term evolution of PPDs under different model variations. Diagnosing angular momentum transport and mass loss, we outline the fundamental framework for the transition from Class 0/I to Class II disks and examine how stellar irradiation and MHD wind introduce additional complexities to this baseline scenario. 

\subsection{Two Primary Models}

We here focus on two primary suites of models that bracket different assumptions about the large-scale vertical magnetic flux: the fiducial model with a prescribed vertical magnetic field strength (\textbf{Bz\_10mG}), and a contrasting model with a constant plasma $\beta_z$ (\textbf{Betaz\_3e4}). These serve as the two reference cases for our study, while additional minor variations—such as different floor values $\beta_{z,\mathrm{min}}$ in the \textbf{Bz\_10mG} model—are introduced later in Section~\ref{sec:discussion} to assess the robustness of our results.

\vspace{0.5em}
\noindent The two primary models are defined as follows:
\begin{itemize}
    \item \textbf{Bz\_10mG} (Fiducial): the uniform-$B_z$ prescription, Equation~(\ref{eq:bz}), together with the full set of physical processes described in Section~\ref{sec:modelsetup}.
    \item \textbf{Betaz\_3e4}: identical to the fiducial model except for the pressure-scaled (constant-$\beta_z$) field, Equation~(\ref{eq:bz2}), with $\beta_z = 3\times10^4$ at all radii.
\end{itemize}

These two prescriptions imply qualitatively different radial distributions of magnetic field. In the \textbf{Bz\_10mG} model, a spatially uniform $B_z$ becomes dynamically stronger toward large radii as the gas density decreases, so the outer disk is especially susceptible to MHD wind-driven processes. In the \textbf{Betaz\_3e4} model, by contrast, $B_z$ scales with the local gas pressure. This produces a much weaker absolute magnetic field in the outer disk, but a relatively stronger field in the inner disk. As we will show, this dichotomy fundamentally alters the disk evolution.

\subsection{Overview}\label{sec:overview}

\begin{figure*}[ht!]
\plotone{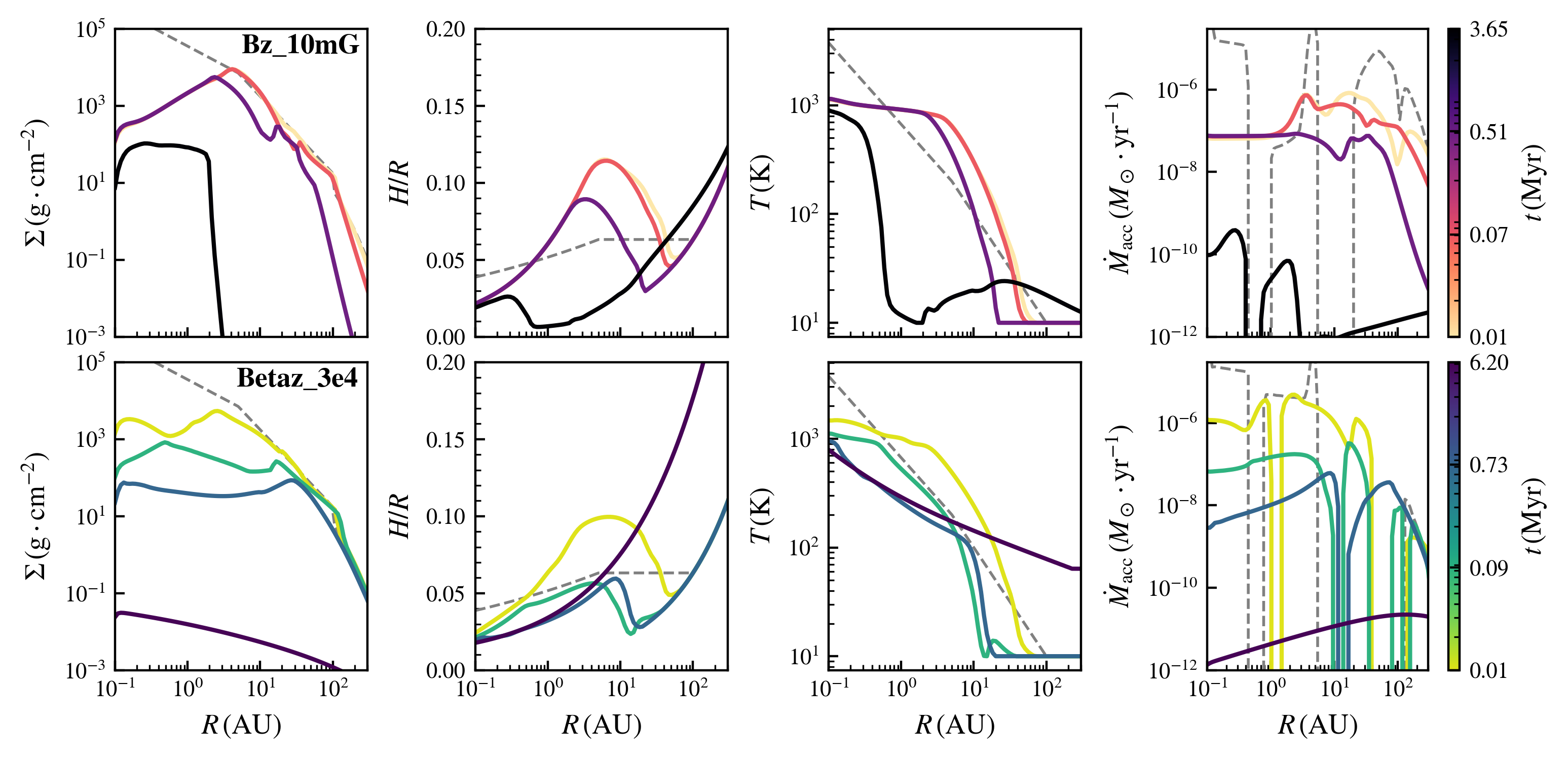}
\caption{Profiles of disk surface density $\Sigma$, aspect ratio $H/R$, temperature $T$, and local accretion rate $\dot{M}_{\rm acc}$ (left to right) in the two primary models, namely \textbf{Bz\_10mG} (fiducial) and \textbf{Betaz\_3e4} from top to bottom. Lines with different colors represent different epochs in a logarithmic-uniform manner for each primary model, spanning $\sim 10^{-2}\,\mathrm{Myr}$ (light) to the final plotted snapshots at $3.65\,\mathrm{Myr}$ and $6.20\,\mathrm{Myr}$ for \textbf{Bz\_10mG} and \textbf{Betaz\_3e4}, respectively. The profiles corresponding to the initial conditions are overplotted with the dashed gray lines in each panel. The $\dot{M}_{\rm acc}$ panels show only the positive (inward) accretion rate; the full signed behavior, including regions of net outward transport, is shown in the space-time maps of Figures~\ref{fig:fiducial_alpha} and~\ref{fig:constbetaz_alpha}.
\label{fig:modelvar_impression}}
\end{figure*}

We calculate the disk evolution from the same initial conditions as described in Section \ref{sec:method_numericalsolver} with the two primary models. The \textbf{Bz\_10mG} model is evolved to $3.65\,\mathrm{Myr}$ and the \textbf{Betaz\_3e4} model to $6.2\,\mathrm{Myr}$; these endpoints correspond to absolute disk masses of order $10^{-4}\,M_\odot$. In Figure \ref{fig:modelvar_impression}, we show profiles of disk surface density $\Sigma$, aspect ratio $H/R$, temperature $T$, and local accretion rate $\dot{M}_{\rm acc}$\footnote{The accretion-rate panel shows a diagnostic local radial mass flux computed from the angular momentum transport terms in Equation~(\ref{eq:master_surfden}) as
\begin{equation}\label{eq:mdotacc_diag}
\dot{M}_{\rm acc}(R,t)=
\frac{6\pi}{R\Omega}\partial_R\left(R^2\alpha_{\rm SS}\Sigma c_s^2\right)
+\frac{3\pi\alpha_{\rm DW}\Sigma c_s^2}{\Omega}.
\end{equation}
Positive values indicate accretion, while negative values indicate outward spreading.} in the two primary models across different epochs. 

To provide a clear global picture of the disk evolution, it is helpful to first divide the disk into three distinct radial segments based on different dominant physical processes \citep{2026arXiv260617150B}. At early stages, both models share the following structural baseline:
\begin{itemize}
    \item The innermost active region ($\sim 0.1-1 \, \mathrm{AU}$): Temperatures exceed $1000 \, \mathrm{K}$ due to a combination of accretion heating and stellar irradiation, sustaining active MRI.
    \item The inner disk ($\sim 1-10 \, \mathrm{AU}$): A dense, optically thick region whose inner edge is directly irradiated and puffed up. This puffed-up rim casts a deep shadow over the GI-active outer disk beyond. In the late phase of the \textbf{Bz\_10mG} model, this zone becomes the long-lived compact reservoir of mass.
    \item The outer disk ($\sim 10$--$100\,\mathrm{AU}$): Initially dense and shielded by the shadow cast from the innermost active region, this part of the disk is marginally gravitationally unstable ($Q \sim 1$) and dominated at early times by GI-driven accretion. Its temperature approaches the ISM background ($\sim 10\,\mathrm{K}$). The fate of this region differs strongly between the two primary models (it is depleted by MHD winds in \textbf{Bz\_10mG} but preserved in \textbf{Betaz\_3e4}).

\end{itemize}
In our model, the disk is truncated beyond $\sim 100\,\mathrm{AU}$. We caution that the actual outer-truncation mechanism in real disks is likely more complex than the standard formalism we adopt here and is not yet well understood \citep[see e.g.,][]{Yang_2021ApJ...922..201Y}. Therefore, we will mainly focus on the evolution of internal disk angular momentum transport and thermal structure, instead of the disk profiles around outer truncation regions. As shown in Figure~\ref{fig:modelvar_impression}, during the disk evolution, their differing magnetic field prescriptions drive vastly different key behaviors across these three regions.

In the \textbf{Bz\_10mG} model, the outer disk experiences runaway mass loss driven by strong MHD disk winds. Because the gas density and thermal pressure naturally decline at larger radii, maintaining a spatially constant vertical magnetic field results in a progressively smaller plasma $\beta_z$ in the outer disk. Consequently, the relatively strong magnetic pressure drives rapid mass loss through MHD disk wind, resulting in a runaway depletion of material within several Myr, with $\beta_z$ held at the imposed floor $\beta_{z,\mathrm{min}}\sim10^2$. In contrast, the inner disk is protected by self-shadowing, as evidenced by the dip in the aspect ratio profiles shown in Figure~\ref{fig:modelvar_impression}. This long-lasting self-shadowing condition limits the strength of the toroidal magnetic field, leading to weaker MHD winds in the inner disk, allowing the compact reservoir to retain most of the system's mass for several Myr. The innermost active region ($R\lesssim 1\,\mathrm{AU}$) remains long-lived, because the spatially constant $B_z$ is dynamically weak relative to the high gas pressure there (i.e., $\beta_z$ is large), so both wind-driven gas removal and MHD wind-driven angular momentum transport are moderate and the inner mass reservoir is not rapidly drained. The upper right panel of Figure \ref{fig:modelvar_impression} shows the accretion-rate profiles in the \textbf{Bz\_10mG} model. At early times, before the outer disk is depleted, the outer disk carries the largest inward mass flux, reaching $\gtrsim 10^{-6}\,M_\odot\,\mathrm{yr}^{-1}$, while the self-shadowed inner reservoir has a lower local accretion rate. At later times, after the outer disk surface density has dropped by orders of magnitude, the depleted outer disk contributes little positive accretion flux, whereas the compact self-shadowed reservoir still maintains a finite but much smaller inward flux, typically $\lesssim10^{-9}\,M_\odot\,\mathrm{yr}^{-1}$ by $3.65\,\mathrm{Myr}$. The detailed mechanisms driving this accretion-rate evolution will be discussed in Section~\ref{sec:am}.

The \textbf{Betaz\_3e4} model, however, preserves its outer disk for $> 1 \, \mathrm{Myr}$ because the magnetic field is weak in low-gas pressure environments, suppressing outer wind-driven mass accretion and depletion. However, its relatively stronger inner magnetic field causes the innermost regions to accrete and deplete much more rapidly. As the surface density in the innermost region drops, it becomes increasingly difficult to sustain MRI and maintain high temperatures. Eventually, the innermost disk ceases to puff up, and the entire disk evolves into a fully flared and extended structure. The accretion-rate profiles support this picture. At early times, while the disk still has high surface density, the constant-$\beta_z$ prescription produces substantial inward flux through the inner and intermediate disk, with $\dot{M}_{\rm acc}$ reaching $\gtrsim10^{-6}\,M_\odot\,\mathrm{yr}^{-1}$ over the inner part of the disk. This reflects the pressure-scaled magnetic field: wind-driven transport is strongest where the disk remains dense and warm, while the weak absolute field in the low-pressure outer disk prevents the rapid outer-disk runaway seen in \textbf{Bz\_10mG}. As the disk surface density and pressure decline, the imposed $B_z$ also weakens, and the positive accretion flux drops to only $\sim10^{-12}$--$10^{-11}\,M_\odot\,\mathrm{yr}^{-1}$ by $6.2\,\mathrm{Myr}$. The detailed mechanisms driving this accretion-rate evolution will be discussed in Section~\ref{sec:am}.

We detail the specific physical mechanisms driving these diverging behaviors in the following subsections: angular momentum transport coupled with disk geometry and thermodynamics (Section~\ref{sec:am}), and the resulting macroscopic disk mass and size observables (Section~\ref{sec:masssize}).

\subsection{Angular Momentum Transport and Thermal Structure}\label{sec:am}

\begin{figure*}[ht!]
\plotone{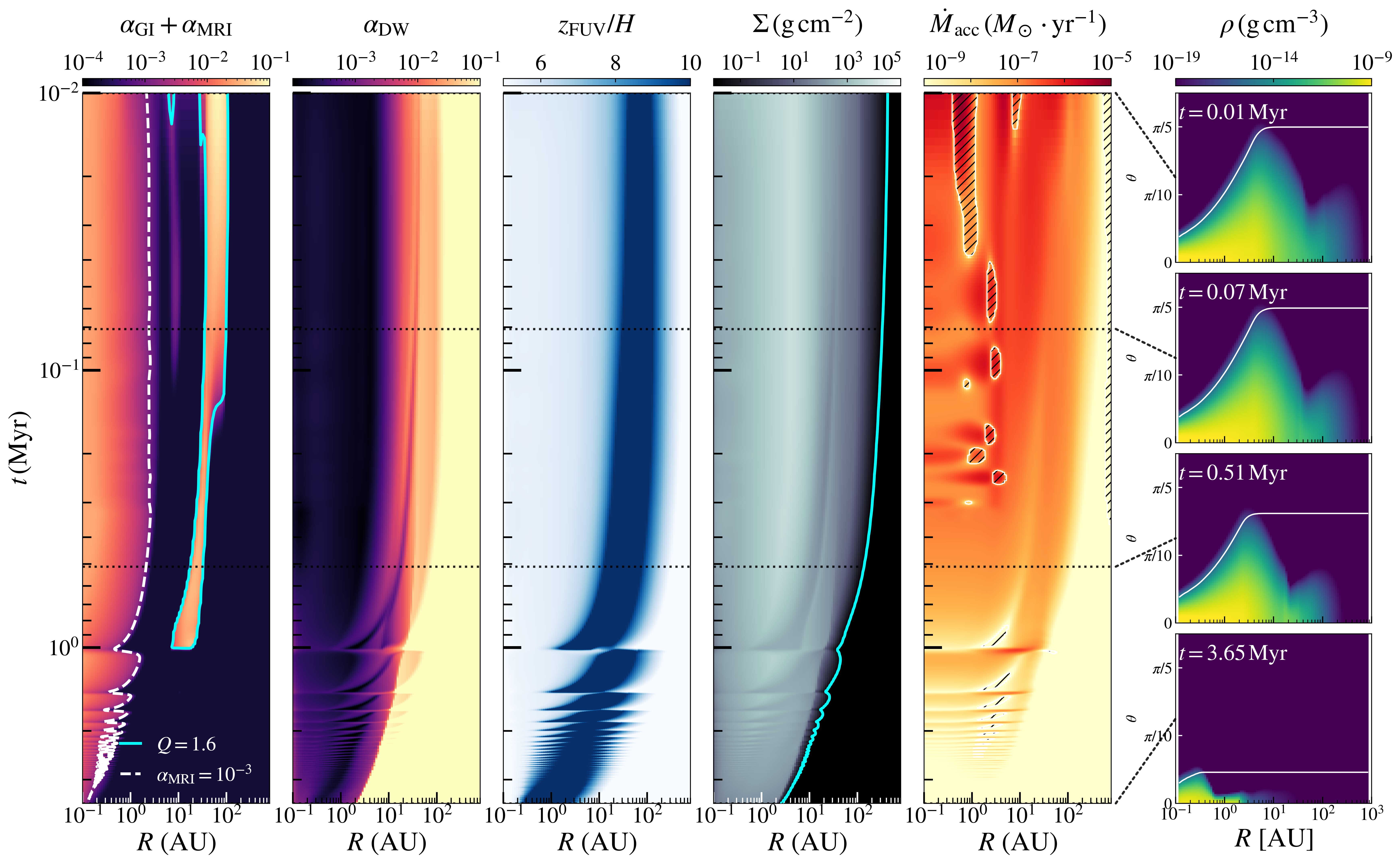}
\caption{Combined diagnostic diagram for the \textbf{Bz\_10mG} model. \textit{Left block:} space-time maps with radius $R$ on the horizontal axis and time $t$ on the logarithmic vertical axis (increasing top-to-bottom from $0.01$ to $3.65\,\mathrm{Myr}$). The five panels show, from left to right, $\alpha_{\rm GI}+\alpha_{\rm MRI}$, $\alpha_{\rm DW}$, $z_{\rm FUV}/H$, $\Sigma$, and local accretion rate $\dot{M}_{\rm acc}$. In the $\dot{M}_{\rm acc}$ panel, the magnitude $|\dot{M}_{\rm acc}|$ is shown on a logarithmic color scale, and hatched regions outlined in white mark net outward (negative) mass flux. The first panel is overlaid with two boundary contours marking the GI- and MRI-active regions: Toomre $Q=1.6$ (cyan solid) and $\alpha_{\rm MRI}=10^{-3}$ (white dashed). The aqua solid contour in the $\Sigma$ panel marks the disk-size threshold $\Sigma=10^{-2}\,\mathrm{g\,cm^{-2}}$. \textit{Right block:} pseudo-2D ($R$--$\theta$) density slices at $t = 0.01$, $0.07$, $0.51$, and $3.65\,\mathrm{Myr}$ (top to bottom), with the FUV ionization-front angle $\theta_{\rm FUV}=\arctan(z_{\rm FUV}/R)$ overlaid as the white solid curve. Dotted lines link each density slice to its corresponding time on the space-time block.
\label{fig:fiducial_alpha}}
\end{figure*}

\begin{figure*}[ht!]
\plotone{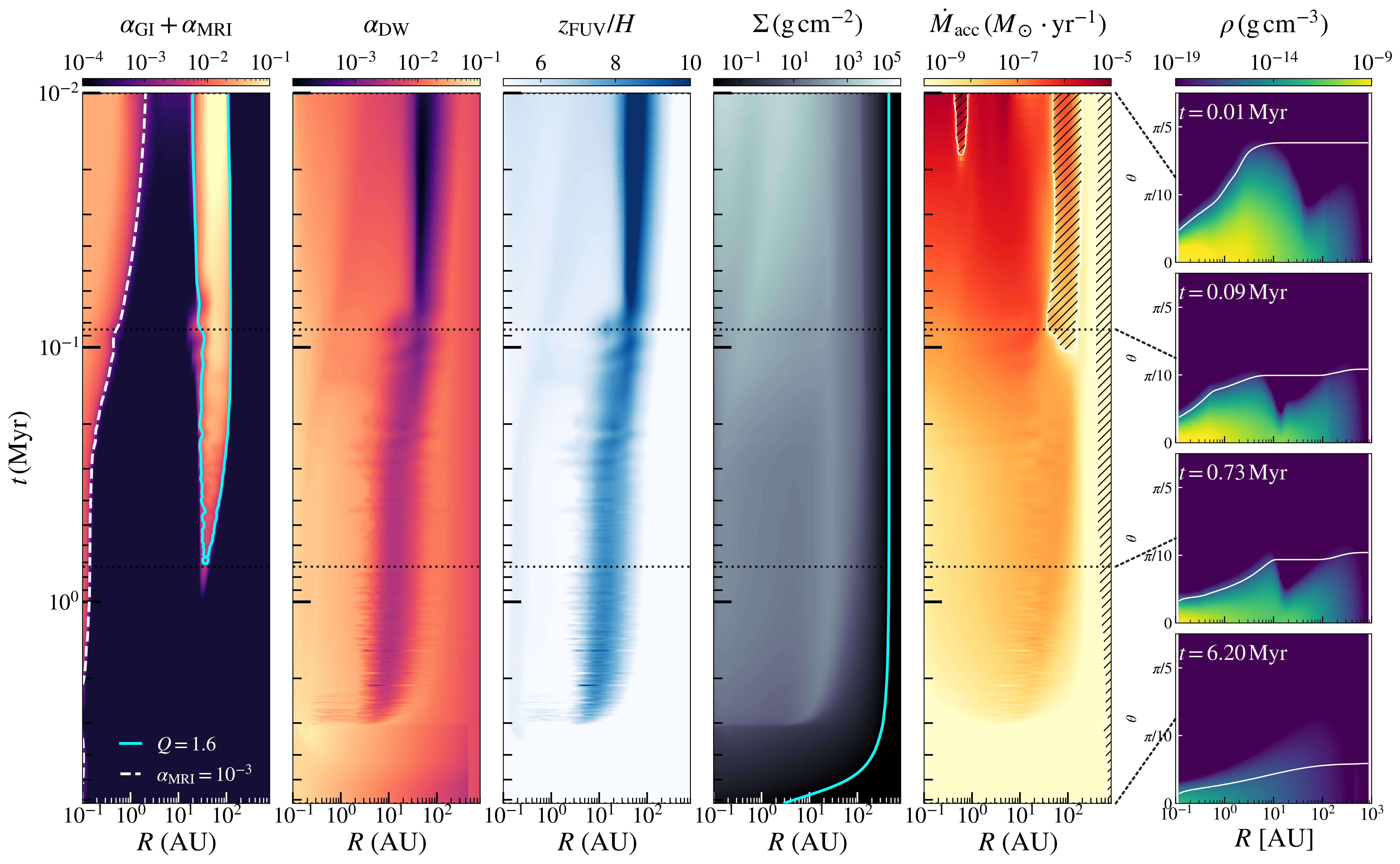}
\caption{Combined diagnostic diagram for the \textbf{Betaz\_3e4} model, in the same layout as Figure~\ref{fig:fiducial_alpha}. The left block shows space-time maps of $\alpha_{\rm GI}+\alpha_{\rm MRI}$, $\alpha_{\rm DW}$, $z_{\rm FUV}/H$, $\Sigma$, and local accretion rate $\dot{M}_{\rm acc}$ (left to right), with the Toomre $Q=1.6$ (cyan solid) and $\alpha_{\rm MRI}=10^{-3}$ (white dashed) boundary contours overlaid on the first panel. The aqua solid contour in the $\Sigma$ panel marks the disk-size threshold $\Sigma=10^{-2}\,\mathrm{g\,cm^{-2}}$. The right block shows pseudo-2D ($R$--$\theta$) density slices at $t = 0.01$, $0.09$, $0.73$, and $6.20\,\mathrm{Myr}$, with the FUV ionization-front angle $\theta_{\rm FUV}$ overlaid as the white solid curve, and dotted lines linking each slice to its corresponding time on the space-time block.
\label{fig:constbetaz_alpha}}
\end{figure*}

We now turn to the underlying physics driving the macroscopic evolution described in Section~\ref{sec:overview}. Because angular momentum transport, irradiation geometry, and thermal structure are intrinsically coupled in this model---the locations of the GI-, MHD wind- and MRI-active zones depend on the temperature and irradiation geometry, which are in turn set by the dissipation profile and by where stellar photons are absorbed---we discuss them together here, treating the two primary models as case studies in the two sub-subsections below.
Figure~\ref{fig:fiducial_alpha} and Figure~\ref{fig:constbetaz_alpha} present space-time maps of various angular momentum transport mechanisms in the two primary models, respectively in their left blocks: the GI-induced effective viscosity $\alpha_{\rm GI}$ and MRI-induced effective viscosity $\alpha_{\rm MRI}$ (first), the wind-torque parameter $\alpha_{\rm DW}$ (second), the dimensionless ionization-front height $z_{\rm FUV}/H$ (third), surface density $\Sigma$ (fourth), and local accretion rate $\dot{M}_{\rm acc}$ (fifth, from left to right).
The right block further shows the corresponding ``pseudo-2D'' ($R$--$\theta$) density distribution at selected epochs (the vertical density profile is assumed Gaussian throughout, hence ``pseudo-2D''), with the FUV-ionization surface overlaid as the white solid curve. We emphasize that $z_{\rm FUV}/H$ is used here as a proxy for disk irradiation: self-shadowed regions typically correspond to $z_{\rm FUV}/H\sim10$, while strongly irradiated regions have $z_{\rm FUV}/H\sim5$.

In both primary models, the early evolutionary phase ($\lesssim 0.1\,\mathrm{Myr}$) is characterized by a clear spatial segregation of transport mechanisms that maps onto the three radial zones introduced in Section~\ref{sec:overview}. Within the innermost active region ($R\lesssim 1\,\mathrm{AU}$), MRI-driven turbulence sustains a hot, thermally ionized layer; the temperature of the puffed-up inner rim is set primarily by MRI dissipation, with sub-dominant contributions from stellar irradiation.
The inner disk ($\sim 1$--$10\,\mathrm{AU}$) sits behind this puffed-up rim and is therefore largely shielded from stellar photons; in the absence of irradiation heating, and with GI essentially inactive at these radii, its temperature is set by radiative cooling against weak local dissipation, with radial radiation transport smoothing the transition between the hot inner rim and the cold midplane behind it.
The marginally GI-unstable outer disk sits further out, in the $\sim 10$--$100\,\mathrm{AU}$ range. There, GI self-regulates the disk toward a state of marginal instability ($Q\sim 1$). The resulting gravito-turbulence efficiently transports mass inward, reducing the local surface density $\Sigma$. Consequently, the disk settles into a progressively cooler $Q\sim 1$ state ($Q\propto c_s/\Sigma$) as $\Sigma$ and $T$ decline together along the $Q\sim 1$ track.
After this shared early phase, the two models diverge sharply, and we treat each separately below.

\subsubsection{\textbf{Bz\_10mG} (Fiducial): long-lived shadowed compact disk}\label{sec:am_bz10mg}

We organize the discussion of the fiducial \textbf{Bz\_10mG} model by radial zone---innermost, inner, outer---and close with a synthesis paragraph that ties the angular momentum transport history to the geometric evolution shown in the right column of Figure~\ref{fig:fiducial_alpha}. 

In the innermost active region ($R \lesssim 1\,\mathrm{AU}$), the MRI-active zone persists through most of the $3.65\,\mathrm{Myr}$ calculation, and begins to weaken when the local positive accretion flux drops to $\lesssim10^{-8}\,M_\odot\,\mathrm{yr}^{-1}$. The MRI activity is sustained primarily by local MRI viscous heating with sub-dominant contributions from stellar irradiation. This is evidenced by the long-lived high-$\alpha_{\rm MRI}$ region in Figure~\ref{fig:fiducial_alpha}: the MRI-sustained high temperature ($T \gtrsim 1000\,\mathrm{K}$) keeps the rim puffed up and casts the deep shadow that shields the disk further out. The MHD wind plays only a secondary role here ($\alpha_{\rm DW}\sim 10^{-3}$): although $z_\mathrm{FUV}/H \sim 5$ indicates strong irradiation and thus a large $B_\phi/B_z$ ratio (Equation~\ref{eq:bphi}), the constant-$B_z$ assumption keeps the vertical magnetic field weak relative to the immense local gas pressure. Because the innermost region remains hot and has large $Q$, GI is never active there.

In addition to local MRI-driven angular momentum transport, the accretion in the innermost active region is regulated by mass supply from larger radii, ultimately associated with the GI-active outer disk at early times. This supply helps maintain the surface density and optical depth needed for MRI heating to keep the gas thermally ionized, and the positive local accretion flux remains of order $\gtrsim10^{-7}\,M_\odot\,\mathrm{yr}^{-1}$ during the early phase.

Soon after GI shuts down in the outer disk around $\sim 1\,\mathrm{Myr}$ (see below), episodic variations in $\alpha_{\rm MRI}$ emerge in this zone: the reduced inflow lowers the innermost surface density and optical depth needed to sustain MRI activity; later mass accumulation can raise the optical depth and temperature again, reactivating the MRI locally. This cycle repeats several times before the inner disk settles into a new quasi-steady state with a reduced MRI-active region, and the surface-density and local accretion-rate panels of the same figure make the episodic structure visible inside $R \sim 1\,\mathrm{AU}$. The $\dot{M}_{\rm acc}$ map shows that these episodes involve local redistribution of mass: positive inward-flow bands are interleaved with weak or outward-flowing regions as the MRI-active zone varies. The resulting cyclic accretion is conceptually reminiscent of FUor-type outburst behavior \citep[e.g.,][]{Bell_1994ApJ,Zhu_2010ApJ}, although the $\sim0.1\,\mathrm{Myr}$ recurrence timescale in our model is longer than the $\sim10^{-3}$--$10^{-2}\,\mathrm{Myr}$ recurrence timescales often inferred for FUor-like episodic accretion \citep[see][for review]{Audard_2014prpl.conf..387A}. However, we caution that the recurrence time and amplitude of these episodes are sensitive to our prescription of magnetic flux evolution.

The mass involved in these episodes is nevertheless small. By the onset of the episodic phase, about $90\%$ of the initial disk mass has already been lost through accretion onto the star and MHD wind depletion. Over the remaining evolution, the episodic accretion onto the star amounts to only a few percent of the initial disk mass, while the inner disk is drained by wind-driven accretion and depletion. Although the episodes dominate the variability of the stellar accretion rate in this phase, the decline of the remaining disk mass and the lifetime of the compact reservoir are governed by the wind-driven drainage of the inner disk.

The massive inner disk ($\sim 1$--$10\,\mathrm{AU}$) acts as the long-lived shielded mass reservoir of the disk. After the outer disk begins to drain, the puffed-up innermost active region keeps this inner disk self-shadowed, with $z_\mathrm{FUV}/H > 10$ from $\sim1\,\mathrm{Myr}$ onward. This weakens the wind torque ($\alpha_{\rm DW}\lesssim10^{-3}$ in the shadowed region), and the local accretion-rate panel shows that $\dot{M}_{\rm acc} \lesssim 10^{-8} M_\odot\,\mathrm{yr}^{-1}$ through most of the reservoir, much smaller than the early outer-disk inflow ($\gtrsim 10^{-6} M_\odot\,\mathrm{yr}^{-1}$). GI survives for $\sim 1 \,\mathrm{Myr}$ here: once the outer disk no longer supplies mass inward, $\Sigma$ declines slowly, and because the gas is already near the $\sim10\,\mathrm{K}$ floor and cannot cool further to maintain $Q\sim1$, $Q$ rises above unity and GI shuts down. The outer part of this zone ($\sim5$--$10\,\mathrm{AU}$) is gradually exposed as the irradiation front moves inward, but the self-shadowed portion remains dense. Even after the outer disk no longer supplies a strong inflow, both transport channels remain weak, so the shadowed part of the inner disk evolves only slowly: the surface-density panel of Figure~\ref{fig:fiducial_alpha} confirms that $\Sigma$ in this dense region remains $\sim 10^{2}$--$10^{4}\,\mathrm{g\,cm^{-2}}$ throughout the $3.65\,\mathrm{Myr}$ calculation, leaving a compact, optically thick, cold reservoir within the shadow.

The outer disk ($\sim 10$--$100\,\mathrm{AU}$) is where the GI-to-wind transport transition plays out, and where the most rapid mass depletion and accretion occur. At early times this region is marginally GI-unstable ($Q\sim1$) and shielded by the inner rim's shadow. GI self-regulation drives $\Sigma$ and $T$ together along the $Q\sim1$ track until the gas reaches the $\sim10\,\mathrm{K}$ floor; after that, $\alpha_{\rm GI}$ declines and wind-driven transport becomes dominant. As $\beta_z$ in the depleted outer disk falls to the imposed floor $\beta_{z,\mathrm{min}}\sim10^2$, the wind-torque parameter rises to $\alpha_{\rm DW}\sim10^{-1}$, and the combination of wind-driven accretion and wind mass loading drains the outer disk on a Myr timescale. In the meantime, the fast wind-driven accretion supplies mass inward, thus making the GI-active zone drift inward within the first Myr. After $\sim 1\,\mathrm{Myr}$, as shown in the $\dot{M}_{\rm acc}$ panel of Figure~\ref{fig:fiducial_alpha}, the initially large inward flux outside $\sim10\,\mathrm{AU}$ fades rapidly as $\Sigma$ is exhausted, though $\alpha_{\rm DW}$ in the depleted outer disk remains high. 

The interplay between these three zones ties them together through a self-reinforcing feedback loop, and the resulting geometric evolution is most directly visible in the density slices in the right column of Figure~\ref{fig:fiducial_alpha}. MRI dissipation puffs up the innermost region, which casts a deep shadow over the inner and outer disk. The shadow keeps the inner disk cool and $z_\mathrm{FUV}/H$ high, simultaneously weakening the MHD wind in the shadowed inner disk. The resulting weak wind preserves the dense, shielded gas in the inner disk, and the inner disk in turn continues to feed the puffed-up innermost MRI zone. The four density slices show this loop in action. At $t = 0.01\,\mathrm{Myr}$ the disk is still extended, with appreciable midplane density out to several tens of AU. By $t = 0.07\,\mathrm{Myr}$ the outer disk has begun to clear, and by $t = 0.51\,\mathrm{Myr}$ the dense gas is concentrated in the compact inner disk behind the puffed inner rim. By $t = 3.65\,\mathrm{Myr}$, the outer disk is completely depleted, but the compact, optically thick shadowed reservoir remains. This long-lived compact reservoir produces the distinctive observable mass and size evolution discussed in Section~\ref{sec:masssize}.

\subsubsection{\texorpdfstring{\textbf{Betaz\_3e4}}{Betaz\_3e4}: early flared transition}\label{sec:am_betaz5}

The \textbf{Betaz\_3e4} model exhibits a markedly different transport history due to its alternative magnetic-field profile: under the constant-$\beta_z$ prescription, $B_z \propto \sqrt{\rho c_s^2}$ scales with the local gas pressure, so the field is strongest where the gas is densest (the inner disk) and weakest where it is most diffuse (the outer disk), reversing the radial profile of $\alpha_{\rm DW}$ relative to \textbf{Bz\_10mG}. We now discuss the per-zone physics as in Section~\ref{sec:am_bz10mg}---innermost, inner, outer---and close with a synthesis paragraph that ties the angular momentum transport history to the geometric evolution shown in the right column of Figure~\ref{fig:constbetaz_alpha}.

In the innermost active region ($R \lesssim 1\,\mathrm{AU}$), the pressure-scaled magnetic field in \textbf{Betaz\_3e4} produces much stronger wind-driven transport ($\alpha_\mathrm{DW}\sim10^{-2}$) than in \textbf{Bz\_10mG}, with early inward flux reaching $\sim10^{-6}$--$10^{-5}\,M_\odot\,\mathrm{yr}^{-1}$ through the joint effect of MHD wind and MRI. As the innermost surface density drops due to the rapid accretion, the disk can no longer maintain the thermally ionized state needed for sustained MRI activity: $\alpha_{\rm MRI}$ declines from its initial $\sim10^{-2}$ to $\lesssim10^{-3}$ within the first $\sim1\,\mathrm{Myr}$ and continues to decay thereafter. This weakening occurs as the local accretion rate falls to $\lesssim10^{-8}\,M_\odot\,\mathrm{yr}^{-1}$, consistent with the findings in the radiative (non-ideal) MHD simulations (e.g., Wang et al. in prep.). Once MRI heating weakens, the rim cools and ceases to puff up, and the deep shadow it had cast over the inner and outer disk is lost---the central event around which the rest of the model's evolution is organized.

The inner disk ($\sim1$--$10\,\mathrm{AU}$) is exposed to direct stellar irradiation once the innermost rim collapses by $\sim1\,\mathrm{Myr}$. As shown in the $z_{\rm FUV}/H$ panel of Figure~\ref{fig:constbetaz_alpha}, the inner rim never establishes a long-lasting shadow over this region, so irradiation rather than local dissipation sets the thermal balance over a much broader radial range than in \textbf{Bz\_10mG}. GI remains essentially inactive here because the surface density does not approach the $Q\sim1$ threshold while the irradiation-set temperature stays well above the $\sim10\,\mathrm{K}$ floor. Instead, the unshielded inner disk hosts a wind-dominated regime with $\alpha_{\rm DW}\gtrsim10^{-2}$. The surface-density and accretion-rate panels of Figure~\ref{fig:constbetaz_alpha} show the consequence: $\Sigma$ declines more uniformly than in \textbf{Bz\_10mG}, while positive inward mass flux is present across much of the $\sim1$--$10\,\mathrm{AU}$ region after the innermost rim collapses.

The outer disk ($\sim10$--$100\,\mathrm{AU}$) is preserved on Myr timescales because the constant-$\beta_z$ prescription gives this low-pressure region a much weaker $B_z$ than in \textbf{Bz\_10mG}, lengthening the wind-driven accretion and depletion timescale. As a result, within the first Myr, the GI-active zone in the outer disk (see the first panel of Figure~\ref{fig:constbetaz_alpha}) remains at radii beyond $\sim10\,\mathrm{AU}$, in contrast with \textbf{Bz\_10mG}, where rapid outer-disk depletion and accretion drive the GI-active region inward and eventually remove it. Within this GI-active outer region, GI-driven transport dominates at early times, with $\alpha_{\rm GI}\sim10^{-2}$--$10^{-1}$, while the MHD wind is sub-dominant with $\alpha_{\rm DW}\sim 10^{-3}$--$10^{-2}$ across the entire outer disk. As in the \textbf{Bz\_10mG} model, the MHD wind becomes the dominant angular momentum transport channel after GI weakens. However, the imposed constant-$\beta_z$ field leaves the outer disk evolving more slowly than in \textbf{Bz\_10mG}; correspondingly, the $\dot{M}_{\rm acc}$ map shows that positive inward accretion persists across the outer disk without the rapid surface-density collapse seen in \textbf{Bz\_10mG}. The surface-density panel of Figure~\ref{fig:constbetaz_alpha} confirms that $\Sigma$ in the outer disk is the most slowly evolving part of the model, decreasing only modestly over most of the $6.2\,\mathrm{Myr}$ calculation.

The interplay between thermodynamics and angular momentum transport in \textbf{Betaz\_3e4} therefore unfolds without the self-reinforcing shadowed-reservoir feedback found in \textbf{Bz\_10mG}, and the resulting geometric evolution is most directly visible in the density slices in the right column of Figure~\ref{fig:constbetaz_alpha}. The stronger wind-driven transport in the dense innermost disk rapidly lowers the surface density, so MRI heating can no longer sustain a hot, puffed-up rim. The rim collapses, the shadow is lost, and stellar photons reach the inner and outer disk. GI fades as those regions warm, leaving MHD winds as the dominant remaining angular momentum transport channel. The four density slices show this sequence in action. At $t = 0.01\,\mathrm{Myr}$ the configuration is similar to that of \textbf{Bz\_10mG}, with the inner rim still puffed and shadowing the disk further out. By $t = 0.09\,\mathrm{Myr}$ the rim has begun to thin, and by $t = 0.73\,\mathrm{Myr}$ the rim has collapsed enough to expose most of the disk to direct stellar irradiation. At $t = 6.20\,\mathrm{Myr}$ the outer disk remains largely preserved and has a much wider radial extent than its \textbf{Bz\_10mG} counterpart near its final snapshot. This globally flared, irradiation-heated, wind-dominated end state produces macroscopic observational signatures that differ qualitatively from the \textbf{Bz\_10mG} reservoir-dominated configuration, as we discuss in Section~\ref{sec:masssize}.

\subsection{Observables}\label{sec:masssize}

\begin{figure*}[ht!]
\plotone{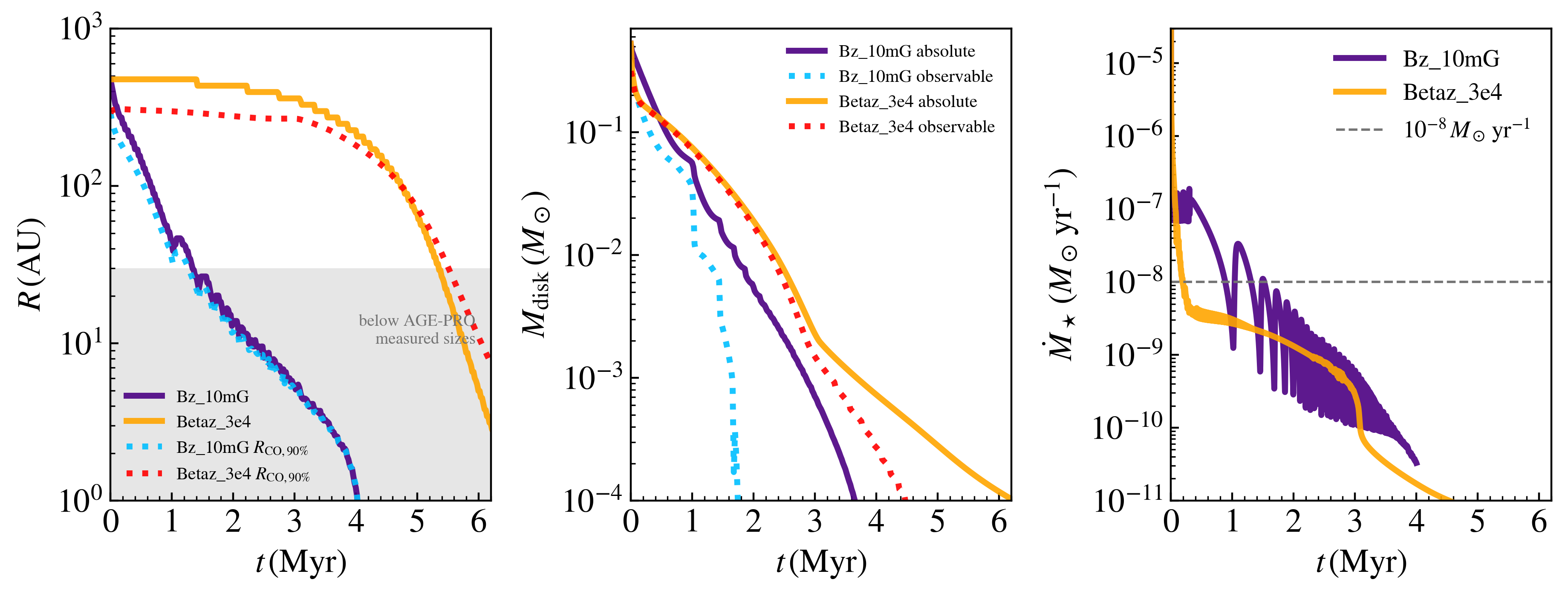}
\caption{The evolution of disk size (left), disk mass $M_\mathrm{disk}$ (middle), and stellar accretion-rate diagnostic $\dot{M}_\star$ (right) for the two primary models. The purple and orange solid lines correspond to \textbf{Bz\_10mG} and \textbf{Betaz\_3e4}, respectively. In the left panel, solid lines show the truncation radius $R_\mathrm{disk}$ and dotted lines the CO emission radius $R_\mathrm{CO,90\%}$; the shaded band marks radii below the smallest gas disk sizes measured by AGE-PRO. In the middle panel, the observable disk mass $M_\mathrm{disk,obs}$ is additionally shown as dotted cyan and red lines for \textbf{Bz\_10mG} and \textbf{Betaz\_3e4}, respectively. In the right panel, $\dot{M}_\star$ is estimated from the inward mass flux through the innermost grid zone, and the dashed horizontal line marks $10^{-8}\,M_\odot\,\mathrm{yr}^{-1}$.
\label{fig:modelvar_disksizemass}}
\end{figure*}

The physical and thermodynamic processes described above ultimately govern the macroscopic, observable properties of the PPDs. In Figure~\ref{fig:modelvar_disksizemass}, we quantitatively show the evolution of disk size, disk mass, and stellar accretion rate for both primary models. We define the model disk size $R_\mathrm{disk}$ as the truncation radius where $\Sigma$ starts to drop below $1\times10^{-2}\,\mathrm{g\,cm^{-2}}$. To compare with gas disk sizes measured from CO emission, we further estimate $R_\mathrm{CO,90\%}$, the radius enclosing $90\%$ of the $^{12}$CO flux, following \citet{Tabone_2025ApJ...989....7T}. We locate $R_\mathrm{CO,90\%}$ at the radius where the gas column density drops below $N_\mathrm{gas}=3.7\times10^{21}\,\delta_\mathrm{C}^{-1}(M_\mathrm{disk,abs}/M_\odot)^{0.34}\,\mathrm{cm^{-2}}$, a criterion calibrated against thermochemical disk models \citep{Trapman_2023ApJ...954...41T}, taking $N_\mathrm{gas}=\Sigma/(2\mu m_\mathrm{H})$. This threshold marks the column below which CO no longer self-shields against photodissociation, and its dependence on disk mass accounts for the larger fraction of CO frozen onto grains in colder, more massive disks. We adopt the gas-phase carbon depletion factor $\delta_\mathrm{C}=0.2$ used by \citet{Tabone_2025ApJ...989....7T} for the Lupus and Upper~Sco samples. We further define an observable disk mass, $M_\mathrm{disk,obs}=\int_{10\,\mathrm{AU}}^{R_\mathrm{CO,90\%}}\Sigma\,2\pi R\,\mathrm{d}R$, as the gas mass accessible to such a measurement, as distinct from the absolute disk mass $M_\mathrm{disk,abs}$ integrated over the full radial domain. The outer limit is the extent of the CO emission. The inner limit is adopted as $10\,\mathrm{AU}$, within which the gas is poorly traced by ALMA \citep[e.g.][]{Tabone_2025ApJ...989....7T}. 

The main evolutionary outcomes for the macroscopic observable disk properties are as follows:

\begin{itemize}
    \item \textbf{Evolution of Disk Size:} The disk size in the \textbf{Bz\_10mG} model shrinks by over one order of magnitude within the first few Myr and reaches a compact ($\lesssim 5\,\mathrm{AU}$) state by the end of the plotted evolution. In contrast, the \textbf{Betaz\_3e4} model preserves a disk larger than $100\,\mathrm{AU}$ for several Myr before undergoing rapid late contraction.
    \item \textbf{Evolution of Total Disk Mass:} In the \textbf{Bz\_10mG} model, the absolute disk mass drops rapidly during the first Myr and then continues declining toward $\sim10^{-4}\,M_\odot$ by $3.65\,\mathrm{Myr}$. The absolute disk mass in the \textbf{Betaz\_3e4} model decreases more gradually at early times, remaining above the \textbf{Bz\_10mG} track for most of the evolution before its final rapid depletion.
    \item \textbf{Observable vs. Absolute Disk Mass:} The relation between $M_\mathrm{disk,abs}$ and $M_\mathrm{disk,obs}$ depends on the disk's structural evolution. In the \textbf{Betaz\_3e4} model, the two track each other to within a factor of two for most of the evolution, because the disk remains extended and the CO emission covers nearly the whole mass distribution. In the \textbf{Bz\_10mG} model, the CO emission radius contracts as the outer disk is dispersed, and the mass that remains is concentrated at radii where CO no longer traces it. After $\sim1.6\,\mathrm{Myr}$, $M_\mathrm{disk,obs}$ falls more than an order of magnitude below $M_\mathrm{disk,abs}$.
    \item \textbf{Evolution of Stellar Accretion Rate:} The right panel shows the stellar accretion-rate diagnostic $\dot{M}_\star$, estimated from the inward mass flux through the innermost grid zone using the local accretion-rate expression in Equation~(\ref{eq:mdotacc_diag}). Both models begin with high accretion rates, $\dot{M}_\star\gtrsim10^{-5}\,M_\odot\,\mathrm{yr}^{-1}$, but then diverge. In \textbf{Bz\_10mG}, $\dot{M}_\star$ rapidly drops to $\sim10^{-7}\,M_\odot\,\mathrm{yr}^{-1}$ and remains above $\sim10^{-8}\,M_\odot\,\mathrm{yr}^{-1}$, the value commonly quoted for solar-mass T~Tauri stars, for roughly the first Myr, consistent with continued feeding of the innermost active region by the compact reservoir. Thereafter it declines through the episodic phase described in Section~\ref{sec:am_bz10mg}, falling below this value from $\sim1.6\,\mathrm{Myr}$ onward and reaching a few $\times10^{-10}\,M_\odot\,\mathrm{yr}^{-1}$ by $3\,\mathrm{Myr}$. In \textbf{Betaz\_3e4}, $\dot{M}_\star$ declines more steadily and falls below $\sim10^{-8}\,M_\odot\,\mathrm{yr}^{-1}$ already at $\sim0.2\,\mathrm{Myr}$, reflecting the rapid loss of the dense inner disk and the absence of a long-lived dense reservoir. Over most of their lifetimes, both models thus accrete at rates below this reference value; we place these rates in the context of the broad observed distribution in Section~\ref{sec:massizeacc}.

\end{itemize}

\section{DISCUSSION}
\label{sec:discussion}

\subsection{Minor Model Variations}\label{sec:minorvar}

To further test the robustness of our primary models and assess the sensitivity of the results to variations in magnetic field distribution, we modify several parameters in the two baseline models while retaining the core model framework. For the \textbf{Bz\_10mG} model, we consider two additional cases with $\beta_{z,\mathrm{min}} = 10$ (Bz\_10mG\_betamin10) and $\beta_{z,\mathrm{min}} =1000$ (Bz\_10mG\_betamin1000) in Equation~(\ref{eq:bz}), representing values an order of magnitude lower and higher than the primary model, respectively. We also examine a power-law magnetic field configuration given by $B_z = \min(0.03\,\mathrm{G},0.03\,\mathrm{G}\cdot (R/1\,\mathrm{AU})^{-1})$ (Bz\_pwl), keeping all other parameters consistent with the primary \textbf{Bz\_10mG} setup. For the \textbf{Betaz\_3e4} model, we include a minor variation with $\beta_z = 10^4$ in Equation~(\ref{eq:bz2}) (Betaz\_1e4). Each of these four minor model variations is evolved until the total disk mass $M_\mathrm{disk}$ falls below $10^{-4}\,M_\odot$. 

\begin{figure*}[ht!]
\plotone{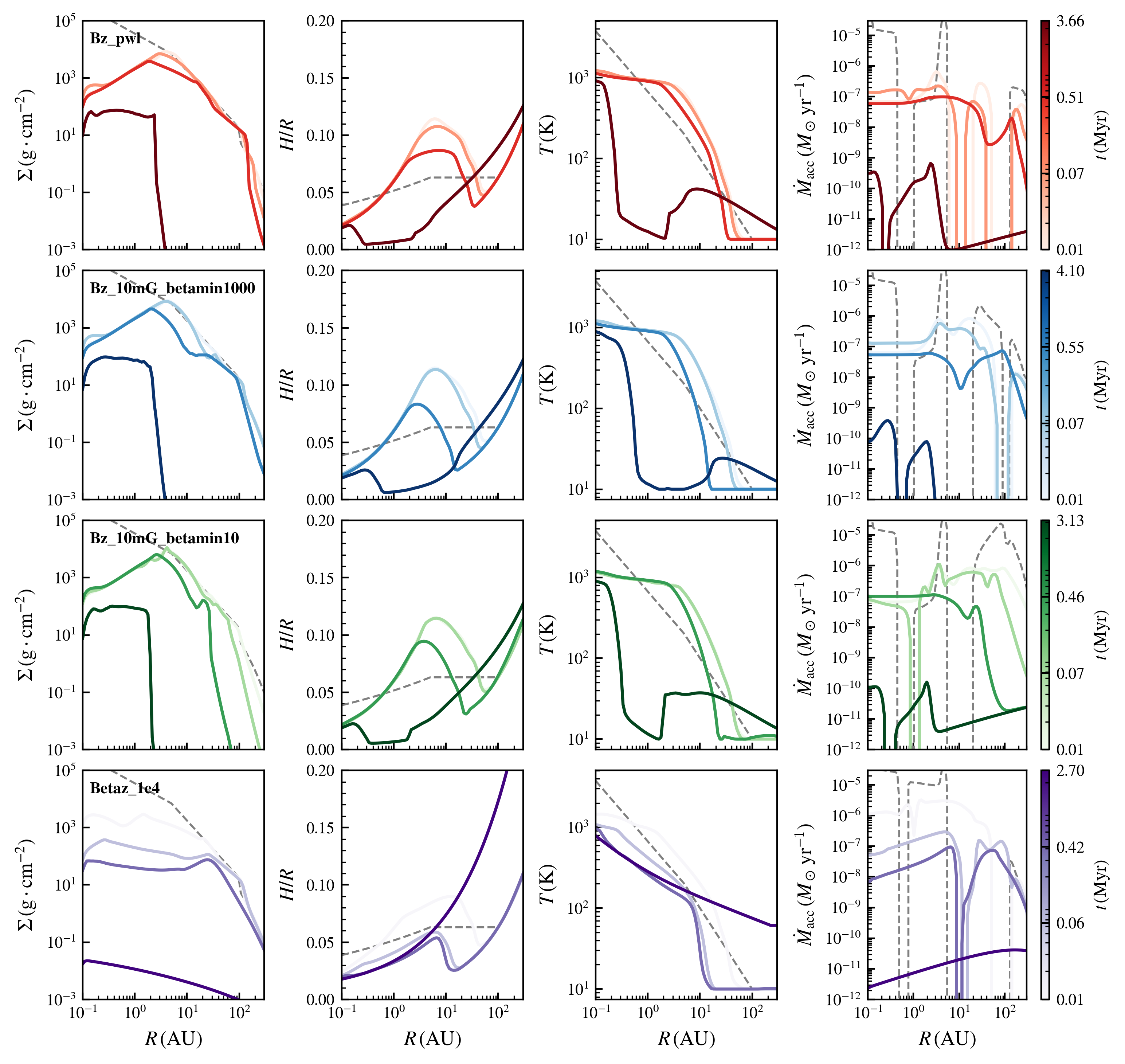}
\caption{Similar to Figure~\ref{fig:modelvar_impression}, but for four minor model variations: Bz\_pwl, Bz\_10mG\_betamin1000, Bz\_10mG\_betamin10, and Betaz\_1e4 from top to bottom. The columns show $\Sigma$, $H/R$, $T$, and local accretion rate $\dot{M}_\mathrm{acc}$ from left to right. Colored curves show logarithmically spaced epochs from $0.01\,\mathrm{Myr}$ to the final plotted epoch of each model, while the gray dashed curves show the initial profiles.
\label{fig:convergence}}
\end{figure*}

As shown in Figure~\ref{fig:convergence}, within the \textbf{Bz\_10mG} family (Bz\_pwl, Bz\_10mG\_betamin1000, and Bz\_10mG\_betamin10), all three variations reproduce the qualitative evolutionary outcome of the fiducial \textbf{Bz\_10mG} model: an early GI-dominated phase in the outer disk, the development of a long-lived self-shadowed bulk disk, and a compact massive reservoir surviving to late times (Figure~\ref{fig:convergence}; cf.\ the top panel of Figure~\ref{fig:modelvar_impression}). Quantitative differences appear mainly in the disk lifetime. The episodic accretion behavior described in Section~\ref{sec:am_bz10mg} appears in all three variations from $\sim1\,\mathrm{Myr}$ onward. In the Bz\_pwl model, the magnetic field is stronger in the dense inner disk compared to the fiducial \textbf{Bz\_10mG} model, leading to enhanced wind-driven accretion and depletion and a final plotted epoch of $3.66\,\mathrm{Myr}$. In the Bz\_10mG\_betamin1000 and Bz\_10mG\_betamin10 models, the modified $\beta_z$ floor primarily affects the low-density outer disk. As a result, their evolutionary tracks are nearly identical to the fiducial model, with slightly longer and shorter final disk lifetimes of $4.10\,\mathrm{Myr}$ and $3.13\,\mathrm{Myr}$, respectively.

Within the \textbf{Betaz\_3e4} family, the Betaz\_1e4 variation---with $\beta_z=10^4$, a factor of three smaller than the fiducial \textbf{Betaz\_3e4}---likewise reproduces the same qualitative end state as its prototype: rapid inner-disk wind depletion, loss of the puffed-up inner rim, and transition to a globally flared, irradiation-dominated configuration. Quantitatively, the stronger MHD wind in Betaz\_1e4 leads to a shorter disk lifetime of $2.70\,\mathrm{Myr}$, again mainly a disk evolution timescale difference rather than a qualitative picture change.

Overall, these tests confirm that the transition from GI-dominated to wind-dominated evolution, as well as self-shadowing, are likely common across minor model variations. For both the \textbf{Bz\_10mG} and \textbf{Betaz\_3e4} model families, the global evolutionary picture remains consistent with their respective prototypes. Parameter variations primarily result in minor structural differences or altered evolutionary timescales.

\subsection{Comparison with Observations}

\subsubsection{Gravitational Instability in Young Disks}

Both our fiducial model (\textbf{Bz\_10mG}) and the \textbf{Betaz\_3e4} model suggest that a marginally gravitationally unstable region outside $\sim 10\,\mathrm{AU}$ can persist for up to $\sim1\,\mathrm{Myr}$. This is broadly consistent with a growing body of evidence that young Class~II disks ($\lesssim$ few Myr) can be sufficiently massive to sit near the threshold of GI, based on non-Keplerian kinematics and spiral-arm morphology \citep{Paneque-Carreno+21, Lodato+23, Speedie_2024Natur, Yoshida_2025NatAs...9.1672Y}, dust-drift constraints on the gas mass \citep{Powell+2019, Williams+2024}, and rare CO isotopologue line emission \citep{Booth+19, BoothIlee20}. These systems are typically older than the epoch at which GI operates in our calculations, and the duration quoted above is an upper limit, since the models contain no sink for the mass processed by GI (fragmentation and planetesimal formation are not followed). Late-stage infall, neglected here (Section~\ref{sec:setup_surfden}), offers an alternative route to massive disks at these ages.

Though all of these probes are sensitive to the outer disk at radii of tens to a hundred AU, some of which are further than where our fiducial model predicts the GI-active region, the qualitative agreement---young disks being massive enough to host GI on tens-of-AU scales---is suggestive.

We caution, however, that this outcome partly reflects our adopted initial condition (where the disk between $\sim 5$ and $100\,\mathrm{AU}$ is initialized near the GI threshold), and the exact radial location of the GI-active part of the disk may also depend on the disk formation process, (proto-)stellar mass, environmental mass replenishment, and large-scale magnetic flux distribution that differ between our fiducial setup and individual observed systems. 

\subsubsection{Compact Disks and the Hidden Mass Reservoir}\label{sec:hiddenmass}

The \textbf{Bz\_10mG} model in particular predicts the survival of a compact ($\lesssim 10\,\mathrm{AU}$), massive, inner disk for several Myr. Compact disks are common in millimeter continuum surveys. \citet{Guerra-Alvarado_2025A&A...696A.232G} report that $\sim 67\%$ of 73 Lupus disks ($\sim1$--$3\,\mathrm{Myr}$) have dust radii below $30\,\mathrm{AU}$, with a median of a few AU, and the AGE-PRO sample gives median dust-disk sizes in Lupus below $15\,\mathrm{AU}$ \citep{Zhang_2025ApJ...989....1Z}. Continuum sizes trace the distribution of millimeter grains and are shaped by their radial drift, while the CO emission, which traces the gas, is on average more extended than the continuum by a factor of a few and correlates with it \citep[e.g.][]{Long_2022ApJ...931....6L}. Compact gas disks are also measured directly. \citet{Ricciardi_2026A&A...710A.112R} find several of the faintest Lupus disks to be consistent with compact, optically thick gas structures, and attribute their weak CO emission to small size and high optical depth.

Despite the existence of compact disks, neither the AGE-PRO sample \citep{Trapman_2025ApJ...989...10T} nor earlier CO surveys spanning $0.5$--$20\,\mathrm{Myr}$ \citep{Long_2022ApJ...931....6L} show a contraction of gas disk size with age, and the compact disks that are observed are generally interpreted as having formed compact \citep{Tabone_2025ApJ...989....7T, Ricciardi_2026A&A...710A.112R}. The rapid contraction of $R_\mathrm{CO,90\%}$ in \textbf{Bz\_10mG} therefore has no counterpart in current measurements. As we discuss in Section~\ref{sec:limitations}, the prescription adopted for the wind torque in our model might overestimate the contraction of the disk size as a function of time.

Compact disks also bear on gas disk mass measurements. As shown in Figure~\ref{fig:modelvar_disksizemass}, in \textbf{Bz\_10mG} the CO emission radius falls quickly, reaching $30\,\mathrm{AU}$---comparable to the smallest gas disk sizes measured by AGE-PRO---at $t\approx1.3\,\mathrm{Myr}$. By this time the majority of the disk mass lies inside $10\,\mathrm{AU}$, where it is poorly traced by CO emission. \citet{Tabone_2025ApJ...989....7T} identify this configuration as a caveat on gas disk mass measurements, noting that the region inside $\simeq10\,\mathrm{AU}$ could hide a large fraction of the mass while remaining poorly traced by ALMA. They further note that such a structure requires a significant jump in the surface density profile and in the efficiency of angular momentum transport. Indeed, as an outcome of the physics described in Section~\ref{sec:am_bz10mg}, our fiducial model develops both. Since the AGE-PRO gas masses are obtained by fitting a grid of thermochemical models in which the gas surface density is a smooth, exponentially tapered power law \citep{Trapman_2025ApJ...989....5T}, the mass concentrated at small radii would not be recovered in the measurement. 

\subsubsection{Self-Shadowed versus Irradiated Disks}

Along with the compact inner disk, the presence of a self-shadowed region during disk evolution is another key feature of our fiducial model (\textbf{Bz\_10mG}). This behavior is qualitatively consistent with several observations: \citet{Flores_2021AJ} capture a disk with a non-monotonic radial temperature profile, showing a transition corresponding to a drop in disk brightness, which aligns with our reduced surface density when the disk becomes re-irradiated. Similarly, IM Lup displays a transition from a self-shadowed region to an irradiated region \citep{Ueda_2024NatAs}, consistent with the self-shadowing structure predicted by our fiducial model. Moreover, the prominent shadow extending out to $\sim10$--$100\,\mathrm{AU}$ in our fiducial model is qualitatively consistent with the shadows seen in scattered-light imaging \citep[e.g.][]{Garufi_2017A&A...603A..21G, Avenhaus_2018ApJ...863...44A, Garufi_2022A&A...658A.137G}. Such shadow geometry in the first $1\,\mathrm{Myr}$ may also help explain why younger disks---being less exposed to stellar irradiation---tend to appear fainter in scattered light \citep{Garufi_2024}.

This connection can be made more specific in light of recent SPHERE demographics. \citet{Garufi_2022A&A...658A.137G} targeted disks that are faint in scattered light and selected by low far-infrared excess; these disks are typically one to two orders of magnitude fainter than benchmark bright disks, and in systems with ALMA data the detected scattered-light emission often covers only a small part of the mm or gas disk. This is the expected signature of an inner rim placing the outer disk in shadow, and is qualitatively close to the irradiation geometry in \textbf{Bz\_10mG}. The Taurus census of \citet{Garufi_2024} further shows that faint disks dominate the observed Class~II population, including massive but faint systems interpreted as extended self-shadowed disks without large cavities. These systems are the closest observational analogues of the shadowed \textbf{Bz\_10mG} branch. By contrast, bright scattered-light disks in the same census are rare and are preferentially associated with large cavities, strong outer-disk illumination, or environmental/companion perturbations; this morphology is closer to \textbf{Betaz\_3e4}, where the inner rim no longer maintains a global shadow and the disk remains directly irradiated over larger radii. The total-intensity and polarized-light survey of \citet{Ren_2023A&A...680A.114R} also emphasizes that disk morphology and recoverability depend strongly on scattering geometry. We therefore interpret the two primary calculations as bracketing two irradiation pathways: compact or extended self-shadowed disks for \textbf{Bz\_10mG}, and weakly shadowed, directly irradiated, bright/flared disks for \textbf{Betaz\_3e4}, rather than as a single universal evolutionary sequence. Scattered-light images trace (sub)micron grains in the uppermost disk layers, and the appearance of shadows there depends on the scale height of these small grains relative to the height at which stellar photons are absorbed. A quantitative connection between the modeled shadow structure and its appearance in scattered light would require radiative transfer calculations based on our models, which we leave to future work.

\subsubsection{Evolution of Disk Mass, Size, and Accretion Rate}\label{sec:massizeacc}

In terms of the disk evolution in different epochs, the recently released ALMA Large Program AGE-PRO \citep{Zhang_2025ApJ...989....1Z} provides the first systematic measurement of gas disk masses and sizes across a sample of 30 disks in star-forming regions spanning $\sim0.5$--$6\,\mathrm{Myr}$ in age, covering the Ophiuchus, Lupus, and Upper~Sco star-forming regions. This dataset offers a natural observational benchmark for the mass and size evolution predicted by our fiducial model (Figure~\ref{fig:modelvar_disksizemass}). The AGE-PRO median gas disk masses drop by roughly an order of magnitude between Ophiuchus ($\lesssim 1\,\mathrm{Myr}$) and Lupus ($\sim1$--$3\,\mathrm{Myr}$) \citep{Zhang_2025ApJ...989....1Z, Trapman_2025ApJ...989....5T}, which is comparable to the early evolution of the absolute disk mass in our models, where $M_\mathrm{disk,abs}$ falls by roughly an order of magnitude per Myr in the fiducial model and somewhat more slowly in \textbf{Betaz\_3e4}. The absolute disk masses in our models continue to decline at a similar rate at later times, while the observed medians remain comparable between Lupus and Upper~Sco. Several effects bear on this comparison. Our model neglects late infall and streamers as a source of mass replenishment at later epochs (Section~\ref{sec:setup_surfden}). The observed samples, in turn, contain only the disks that have survived and are detected, so the median statistics may be biased. \citet{Trapman_2025ApJ...989....5T} note a significant overlap in stellar age between the Lupus and Upper~Sco samples, with the younger Upper~Sco sources tending to be the more massive ones, which would act to flatten the sequence of medians. Gas masses inferred for compact disks are in addition subject to the effect described in Section~\ref{sec:hiddenmass}. It therefore remains unclear whether the similarity of the Lupus and Upper~Sco medians reflects a genuine slowing of the disk mass evolution. The several-Myr survival times in the two primary models also lie within this observational age range, although the comparison should be interpreted qualitatively because inferred star-forming-region ages and empirical disk lifetimes remain uncertain at the factor-of-few level \citep{2013MNRAS.434..806B}.

The stellar accretion rate ($\dot{M}_\star$, Figure~\ref{fig:modelvar_disksizemass}) provides another observational diagnostic. The initial $\dot{M}_\star\gtrsim10^{-6}\,M_\odot\,\mathrm{yr}^{-1}$ is consistent with the high and highly variable rates inferred for embedded Class~0/I systems \citep[e.g.][]{Dunham_2014prpl.conf..195D,Hartmann_2016ARA&A..54..135H}. At later times, however, $\dot{M}_\star$ in both models drops below the value of $\sim10^{-8}\,M_\odot\,\mathrm{yr}^{-1}$ commonly quoted for solar-mass T~Tauri stars, reaches the low end ($\sim10^{-10}\,M_\odot\,\mathrm{yr}^{-1}$) of the broad range observed for Class~II/T~Tauri disks \citep[e.g.][]{Gullbring_1998ApJ...492..323G,Hartmann_1998ApJ...495..385H,Herczeg_2008ApJ...681..594H,Alcala_2017A&A...600A..20A,Nisini_2018A&A...609A..87N,Manara_2023ASPC} by $\sim3\,\mathrm{Myr}$, and falls below it thereafter. While this late-time shortfall partly stems from our omission of late-stage mass replenishment (Section~\ref{sec:setup_surfden}), the decline rate also depends on the prescription of magnetic flux transport. Because the MHD wind torque scales with the vertical field strength, faster magnetic flux loss or a redistribution of flux away from the outer disk would weaken the wind-driven accretion and depletion, moderating the decline of $\dot{M}_\star$.

\subsection{Implications for planet formation}\label{sec:dustgrowth}

\subsubsection{Dust Growth and Planetesimal Formation}

In our fiducial model (\textbf{Bz\_10mG}), GI already becomes very weak across all radii after $\sim 1\,\mathrm{Myr}$, which translates into a relatively low turbulence. In the meantime, a relatively massive disk survives within $\sim 10\,\mathrm{AU}$ under the protection of self-shadowing, so there is sufficient gas to provide an appreciable dust budget if solids can be retained locally with a one-percent-level dust-to-gas mass ratio. Moreover, within the self-shadowed region the midplane temperature settles near our imposed $\sim 10\,\mathrm{K}$ floor (we note this is a numerical floor here rather than an observational prediction, in principle it can be even lower), corresponding to a low sound speed. Therefore, the self-shadowed regions in the inner disk could be favorable places for dust growth and further planet formation.

To illustrate this point, here we make order-of-magnitude estimates about the dust growth limit according to the turbulence level in the fiducial disk. We can naively calculate dust size corresponding to the turbulent fragmentation barrier as $a_\mathrm{frag}=\frac{2\Sigma}{3\pi\rho_\mathrm{s}\alpha}(\frac{v_\mathrm{frag}}{c_s})^2$ where $\Sigma$ is the gas surface density, $\rho_\mathrm{s}$ is the density of dust grain, $v_\mathrm{frag}$ is the fragmentation threshold velocity and $c_s$ is the sound speed. According to our disk evolution in the primary \textbf{Bz\_10mG} model, in the self-shadowed region $\sim 7\,\mathrm{AU}$ at $\sim 0.9\,\mathrm{Myr}$, the GI becomes weak (Toomre $Q\sim2$) and $\alpha_\mathrm{SS} \sim 10^{-4}$ according to our model. The temperature is $\sim 10.2\, \mathrm{K}$, which translates into $c_s \sim 190\,\mathrm{m\,s^{-1}}$ and the surface density $\Sigma \sim 4.3\times 10^2\,\mathrm{g\,cm^{-2}}$. If we take the value $\rho_\mathrm{s}\sim1\,\mathrm{g\,cm^{-3}}$ \citep{Birnstiel_2018ApJ}, then $a_\mathrm{frag}\sim 0.26\,\mathrm{m}$ for $v_\mathrm{frag}\sim 1\,\mathrm{m\,s^{-1}}$. This corresponds to a Stokes number of ${\rm St}_{\rm frag}\sim 0.09$, which could be sufficient to trigger planetesimal formation by streaming instability \citep{Lim_2024}. Applying the same estimate to \textbf{Betaz\_3e4}, the cold outer disk near $R\sim24\,\mathrm{AU}$ at $t\sim1\,\mathrm{Myr}$ has $T\sim10\,\mathrm{K}$, $c_s\sim188\,\mathrm{m\,s^{-1}}$, and $\Sigma\sim8.1\times10^1\,\mathrm{g\,cm^{-2}}$, giving $a_\mathrm{frag}\sim5\,\mathrm{cm}$ and ${\rm St}_{\rm frag}\sim0.09$. Thus the lower surface density in this model mainly reduces the absolute grain size, while the similar cold temperature and low turbulence lead to a comparable fragmentation-limited Stokes number, in favor of planetesimal formation.

\subsubsection{Substructure Formation}

Another implication of our models concerns substructure formation. In our calculations, surface-density and pressure structures naturally develop at the transitions between different physical zones, e.g., at the outer edge of the MRI-active innermost region (e.g., the dead zone inner boundary), across the shadow boundary where irradiation sets in, and around the GI-active region. These structures arise from the spatial variations of the transport coefficients and from the thermodynamic feedback, rather than from a prescribed radial profile of wind-driven mass loss as in conventional 1D wind models \citep[e.g.][]{Suzuki_2016A&A}.

While the detailed properties of these substructures are unlikely to be quantitatively realistic given our parameterized prescriptions, they demonstrate that once thermodynamics and magnetic flux evolution are incorporated, the disk physics becomes much richer, with substructures emerging at several disk locations. Such locations could act as preferred sites to trap dust \citep[e.g.][]{2012A&A...538A.114P}, promote planetesimal formation \citep[e.g.][]{2005ApJ...620..459Y, 2022ApJ...937L...4X, 2023MNRAS.526...80L}, and stall planet migration \citep[e.g.][]{2006ApJ...642..478M, Cao_2026ApJ}. Because these structures are tied to the transitions between physical regimes, such as the dead-zone inner boundary, the shadow boundary, and the edge of the GI-active region, they are expected to appear at a small number of specific locations in a given disk. The mechanism discussed here may therefore contribute to a subset of the observed substructures.

\subsection{Limitations and Future Directions}\label{sec:limitations}

The primary source of uncertainty in our model lies in the distribution of magnetic fields within protoplanetary disks (PPDs), which remains poorly constrained due to a lack of theoretical understanding and the challenges of direct observational measurements. In our fiducial setup, we adopt a simplified assumption of a spatially uniform vertical magnetic field, subject to a floor on the plasma $\beta_z$. While we explore several alternative magnetic configurations and find that the overall evolutionary trends remain robust across these variations, the true magnetic field structure in real PPDs is likely to be far more complex and diverse. Furthermore, as recently shown by \cite{Yang_2021ApJ...922..201Y}, MHD winds can in principle drive the outer disk to extend outward. The underlying complex dynamics are not easily captured by our parameterized $\alpha_{\rm DW}$ prescription.

While we assume a constant solar luminosity for the central star throughout the disk's evolution, stellar luminosity---particularly in young systems---may vary with accretion rate and other stellar properties in reality. We have tested the impact of a tenfold increase in stellar luminosity and found that our key results remain qualitatively unchanged, though a more systematic exploration of this parameter is planned for future work.

In Section~\ref{sec:dustgrowth}, we roughly estimate the dust growth condition. Here we further note the possibility to incorporate full dust dynamics \citep[e.g.][]{Birnstiel_2010A&A...513A..79B,Birnstiel_2012A&A...539A.148B,Sellek_2020MNRAS, Xu_2023ApJ} and self-consistently describe the temporal evolution of dust content and dust size distributions in our model, as a future extension of our current framework.

\section{CONCLUSIONS}
\label{sec:conclusion}

We have presented a $1+1\mathrm{D}$ global evolution model of PPDs that captures the transition from GI-dominated accretion in the early Class 0/I phase to MHD wind-driven evolution characteristic of the Class II phase. By incorporating radial radiation transport, stellar irradiation, and parameterized angular momentum transport channels—including GI, MRI, and MHD winds—our model offers a coherent framework that connects thermodynamics and accretion physics throughout the disk's lifetime. 

In both primary models (\textbf{Bz\_10mG} and \textbf{Betaz\_3e4}) and their minor-variation families, the early phase ($\lesssim 1\,\mathrm{Myr}$) is set by GI in a cold, self-shadowed outer disk: $\alpha_\mathrm{GI}$ self-regulates the disk toward $Q \sim 1$ by lowering the temperature as the surface density declines, until the temperature reaches the $\sim 10\,\mathrm{K}$ floor and GI quenches locally. The hot innermost disk ($R \lesssim 1\,\mathrm{AU}$) simultaneously sustains MRI-driven turbulence powered chiefly by viscous dissipation, with the MRI-heated rim puffing up and shadowing the disk behind it. At later times ($\gtrsim 1\,\mathrm{Myr}$), as GI fades, MHD winds emerge as the dominant angular momentum transport channel, with gas depletion tied to the adopted lever-arm prescription. The local accretion rate remains radially structured and can even change sign, so a large local transport coefficient does not necessarily imply a large stellar accretion rate.

Beyond these common features, the two model families diverge qualitatively depending on where the wind torque and associated gas depletion are strongest, which is set by the magnetic field prescription. In the \textbf{Bz\_10mG} (fiducial) family, $B_{z}$ is uniform, so the wind torque and associated depletion are strongest in the diffuse outer disk (where $\beta_{z}$ is lowest). We find that:

\begin{itemize}
    \item The outer disk is stripped within $\sim 1\,\mathrm{Myr}$, while the shadow cast by the MRI-heated inner rim protects the inner disk: weak FUV penetration weakens MHD winds there and the low temperature keeps GI alive. As a result, a massive, marginally gravitationally unstable, self-shadowed region survives for $\sim 1\,\mathrm{Myr}$ and feeds the MRI-active innermost disk via GI-driven mass transport.
    \item After GI quenches, the shadow-protected inner disk evolves into a compact ($\lesssim10\,\mathrm{AU}$), cold ($\sim 10\,\mathrm{K}$), low-turbulence ($\alpha_\mathrm{SS}\sim 10^{-4}$), high-density ($\Sigma\gtrsim 300\,\mathrm{g\,cm^{-2}}$) reservoir that survives for several Myr, while the MRI-active zone relaxes to a narrow quasi-steady region at $\sim0.1$--$0.3\,\mathrm{AU}$.
\end{itemize}

These evolutionary features carry several implications. The combination of low turbulence, high surface density, and cold temperature in the shadowed region, sustained over Myr timescales, provides favorable conditions for dust growth and planetesimal formation. Meanwhile, the compact, optically thick inner disk can hide a significant fraction of the total disk mass, leading to an underestimation of the disk mass when inferred from CO emission. This long-lived compact massive disk is qualitatively consistent with the following findings inferred from observations: (i) the interpretation of several faint Lupus disks as compact, optically thick gas structures; (ii) the presence at young ages of disks massive enough to sit near the threshold of gravitational instability; (iii) the reduced scattered-light brightness of younger disks; and (iv) the observed transitions between shadowed and irradiated regions in individual systems. When the size and mass criteria of the AGE-PRO survey are applied to our models, the CO emission radius of \textbf{Bz\_10mG} reaches the smallest gas disk sizes in that survey at $t\approx1.3\,\mathrm{Myr}$, by which time most of the remaining mass lies inside $10\,\mathrm{AU}$. This is the configuration identified as a caveat on gas disk mass measurements \citep{Tabone_2025ApJ...989....7T}, and in our calculations it arises from the interplay between self-shadowing and MHD wind-driven accretion.

In the \textbf{Betaz\_3e4} family, $B_{z}$ scales with the local gas pressure, so the wind torque and associated depletion are instead strongest in the dense inner disk. We find that:
\begin{itemize}
    \item The GI-active outer disk persists for $\lesssim1\,\mathrm{Myr}$ without the clear inward drift of the GI-active zone seen in the \textbf{Bz\_10mG} family, because the weak magnetic field in the low-pressure outer disk limits rapid wind-driven accretion and depletion. Once the innermost rim weakens and the disk becomes more directly irradiated, the outer disk warms and GI gradually fades.
    \item The dense inner disk undergoes stronger wind-driven accretion and depletion than in the \textbf{Bz\_10mG} family. As the surface density and optical depth in the innermost region drop, MRI heating can no longer sustain the puffed-up inner rim, and the shadow-casting structure is gradually lost. The disk then transitions into a globally flared, irradiation-heated configuration in which MHD winds dominate the angular momentum transport and the surface density smoothly depletes over the subsequent evolution.
\end{itemize}

The cold outer disk at $\sim1\,\mathrm{Myr}$ can also provide favorable local conditions for dust growth after GI weakens but before the disk is fully irradiated. In our order-of-magnitude estimate, the lower surface density in this model gives a smaller fragmentation-limited grain size than in \textbf{Bz\_10mG}, but a comparable Stokes number, suggesting that planetesimal formation may still be possible if solids are retained or concentrated locally. Observationally, this branch corresponds more naturally to extended disks with weak shadows or direct irradiation. Because the disk remains extended, the CO emission continues to represent the bulk mass distribution, and the observable disk mass tracks the absolute disk mass more closely than in the \textbf{Bz\_10mG} branch.

Within each model family, minor parameter variations yield qualitatively similar evolution, confirming the robustness of these conclusions, although disk lifetimes and quantitative details remain sensitive to the large-scale magnetic-flux distribution. This bimodal behavior also suggests that the two model pathways may help interpret a broader range of observed PPDs, from compact self-shadowed disks to more extended, directly irradiated disks. At the same time, two discrepancies with current observations remain. The stellar accretion rates in both models reach the low end of the observed Class~II range by $\sim3\,\mathrm{Myr}$ and fall below it thereafter, and the rapid gas disk size contraction along the \textbf{Bz\_10mG} pathway is not reflected in current gas disk size surveys, which show no clear decrease of CO radii with age. The former can be reconciled by incorporating late-stage mass replenishment (Section~\ref{sec:setup_surfden}), and the latter may reflect the adopted prescription of magnetic flux transport (Section~\ref{sec:limitations}).

Taken together, these results point to three broader conclusions concerning PPD evolution. First, \textbf{disk physics is strongly inhomogeneous in both space and time}: local accretion rates and ionization levels vary by orders of magnitude, while the dominant angular momentum transport mechanism and thermal structure change qualitatively across the disk and across evolutionary epochs. Models adopting a single constant $\alpha$---or even a simple superposition of constant $\alpha$'s---cannot capture this behavior and risk missing essential physics. Second, \textbf{thermodynamics and disk geometry play an active rather than a passive role}. Self-shadowing is not merely a consequence of the evolution but provides major dynamical feedback: a shadow simultaneously enables GI by keeping the gas temperature low and weakens MHD winds by limiting FUV penetration and thus magnetic coupling. Whether a self-shadowed massive region persists therefore governs both the dominant transport channel and the dust-growth environment over Myr timescales. Third, \textbf{the transport and distribution of large-scale magnetic flux is the key uncertainty} in disk evolution: different magnetic flux configurations lead to qualitatively different evolutionary pathways, disk lifetimes, and observable properties. Constraining the radial $B_{z}$ profile---both theoretically and observationally---is therefore essential for connecting disk models to observed populations.

\begin{acknowledgments}

Y.N. thanks Jing Yang and Tianhao Li for valuable discussions. This research utilized the \textsc{PYTHON} packages \textsc{NumPy} \citep{harris2020array} and \textsc{SciPy} \citep{2020SciPy-NMeth} for solving partial differential equations and conducting data analysis, and \textsc{Matplotlib} \citep{Hunter:2007} for creating figures. This work is supported by the National Science Foundation of China under grant No. 12325304 and 12233004.

\end{acknowledgments}

\bibliography{manuscript}{}
\bibliographystyle{aasjournalv7}

\end{CJK*}
\end{document}